\documentclass[conference]{IEEEtran}
\IEEEoverridecommandlockouts
\usepackage{cite}
\usepackage{amsmath,amssymb,amsfonts}

\usepackage{booktabs}
\usepackage{siunitx}
\usepackage{balance}

\usepackage{algorithm}
\usepackage{algcompatible} 

\usepackage{color,soul}
\usepackage{enumitem}
\usepackage{graphicx}
\graphicspath{ {./images/} }
\usepackage{textcomp}
\usepackage{xcolor}
\usepackage[export]{adjustbox}
\usepackage{cuted}
\newtheorem{definition}{\textbf{Definition}}

\begin{document}

\title{Energy Loss Prediction in IoT Energy Services
}

\author{\IEEEauthorblockN{Pengwei Yang, Amani Abusafia, Abdallah Lakhdari, and Athman Bouguettaya}
\IEEEauthorblockA{School of Computer Science\\
 The University of Sydney, Australia\\	
pyan8871@uni.sydney.edu.au,\\\{amani.abusafia,abdallah.lakhdari,athman.bouguettaya\}@sydney.edu.au}
}

\maketitle

\begin{abstract}

We propose a novel \textit{Energy Loss Prediction(ELP)} framework that estimates the energy loss in sharing crowdsourced energy services.  Crowdsourcing wireless energy services is a novel and convenient solution to enable the ubiquitous charging of nearby IoT devices. Therefore, capturing the wireless \textit{energy sharing loss} is essential for the successful deployment of efficient energy service composition techniques. We propose \textit{Easeformer}, a novel attention-based algorithm to predict the battery levels of IoT devices in a crowdsourced energy sharing environment. The predicted battery levels are used to estimate the energy loss. A set of experiments were conducted to demonstrate the feasibility and effectiveness of the proposed framework. We conducted extensive experiments on real wireless energy datasets to demonstrate that our framework significantly outperforms existing methods.
\end{abstract}


\section{Introduction}



Internet of things (IoT) is a paradigm that enables everyday objects (i.e., things) to connect to the Internet and exchange data \cite{AtzoriLuigi2010TIoT}. IoT devices, such as smartphones and wearables, typically have augmented capabilities, including sensing, networking, and processing \cite{whitmore2015internet}. Abstracting the capabilities of IoT devices using the \textit{service paradigm} may yield a multitude of novel IoT services \cite{lakhdari2021fairness}\cite{de2011building}. These IoT services may be exchanged between IoT devices as \textit{crowdsourced} IoT services. IoT services are defined by their functional and non-functional attributes \cite{abusafia2022services}. The functional attributes are the tasks performed by an IoT device such as WiFi hotspot access. The non-functional attributes of IoT services are the Quality of Service (QoS) surrounding the delivery of the service, which includes trust and reliability \cite{ba2022multi}. Crowdsourced IoT services refer to the delivery of services from nearby IoT devices \cite{abusafia2022services}.  IoT devices can crowdsource a variety of services, including computing resources \cite{habak2015femto}, energy sharing~\cite{lakhdari2021proactive}\cite{lakhdari2020composing}, and environmental monitoring \cite{kelly2013towards}. For instance, in energy sharing, IoT devices (service providers) may deliver energy \textit{wirelessly} to another nearby device with a low battery (service consumer) \cite{lakhdari2020composing}. This paper focuses on energy services.\looseness=-1

\emph{Energy-as-a-Service (EaaS)}, refers to the wireless delivery of energy among nearby IoT devices \cite{lakhdari2021fairness}\cite{lakhdari2018crowdsourcing}. Energy providers such as smart textiles or solar watches may \textit{harvest} energy from natural resources (e.g., body heat or physical activity) \cite{tran2019wiwear}\cite{li2023activity}\cite{abusafia2022maximizing}. For instance, the PowerWalk kinetic energy harvester produces 10-12 watts of on-the-move power\footnote{bionic-power.com}. The harvested spare energy may be shared with nearby IoT devices as a service. Energy providers may deliver their services using the recently developed wireless charging technologies \cite{yang2023monitoring}\cite{yao2022wireless}\cite{yang2022towards}. For instance, several mobile applications have been developed in recent studies as a first attempt to \textit{enable peer-to-peer wireless energy services over a distance }\cite{yang2023monitoring}\cite{yao2022wireless}\cite{yang2022towards}. These applications allow smartphones to request energy services from nearby smartphones by size, e.g., to be charged by 1000 mAh, or by time, e.g., to be charged for the next 10 minutes. Although the present technology may not offer efficient energy delivery \cite{feng2020advances}, technological advancements are anticipated to facilitate devices to exchange greater quantities of energy \cite{abusafia2022services}\cite{sakai2021towards}. Moreover, several companies, such as Xiaomi (mi.com) and Energous (energous.com), are working on developing technologies to increase both the distance and amount of energy that can be transferred. For instance, Energous developed a technology to enable wireless charging of up to 3 Watts of power within a 5-meter distance\footnote{energous.com}.\looseness=-1

A crowdsourced EaaS ecosystem is a dynamic environment where \emph{providers} and \emph{consumers} gather in \emph{microcells}, such as restaurants. IoT users may share spare energy or request energy from nearby devices. An energy service composition framework has been proposed to manage the allocation of services to requests \cite{abusafia2022services}\cite{lakhdari2020Vision}. The framework assumes that energy requests are fulfilled by an exact amount of advertised energy. However, in reality, consumers receive less energy than expected \cite{yang2023monitoring}\cite{yang2022towards}. Several factors contribute to\textit{ energy loss} during wireless energy transfer, including the type of technology employed, the distance between devices, and the consumption behavior of IoT devices \cite{dhungana2020peer}\cite{lu2015wireless}.

Estimating the \textit{energy loss} is crucial in assessing energy sharing efficiency. Capturing the wireless energy sharing loss is essential for the successful deployment of efficient energy service composition techniques. Energy loss may inform the service composition to optimize the allocation of energy services. For example, choosing a smaller energy service could be better than selecting a larger service due to distance. Furthermore, considering energy loss in the composition ensures meeting consumers' expectations \cite{lakhdari2020composing}. Consequently, selecting services with low energy loss encourages consumer participation in this ecosystem \cite{lakhdari2020Vision}.\looseness=-1





 \begin{figure}[!t]
    \centering
     \setlength{\abovecaptionskip}{-2pt}
    \setlength{\belowcaptionskip}{-25pt}
        \includegraphics[width=0.75\linewidth]{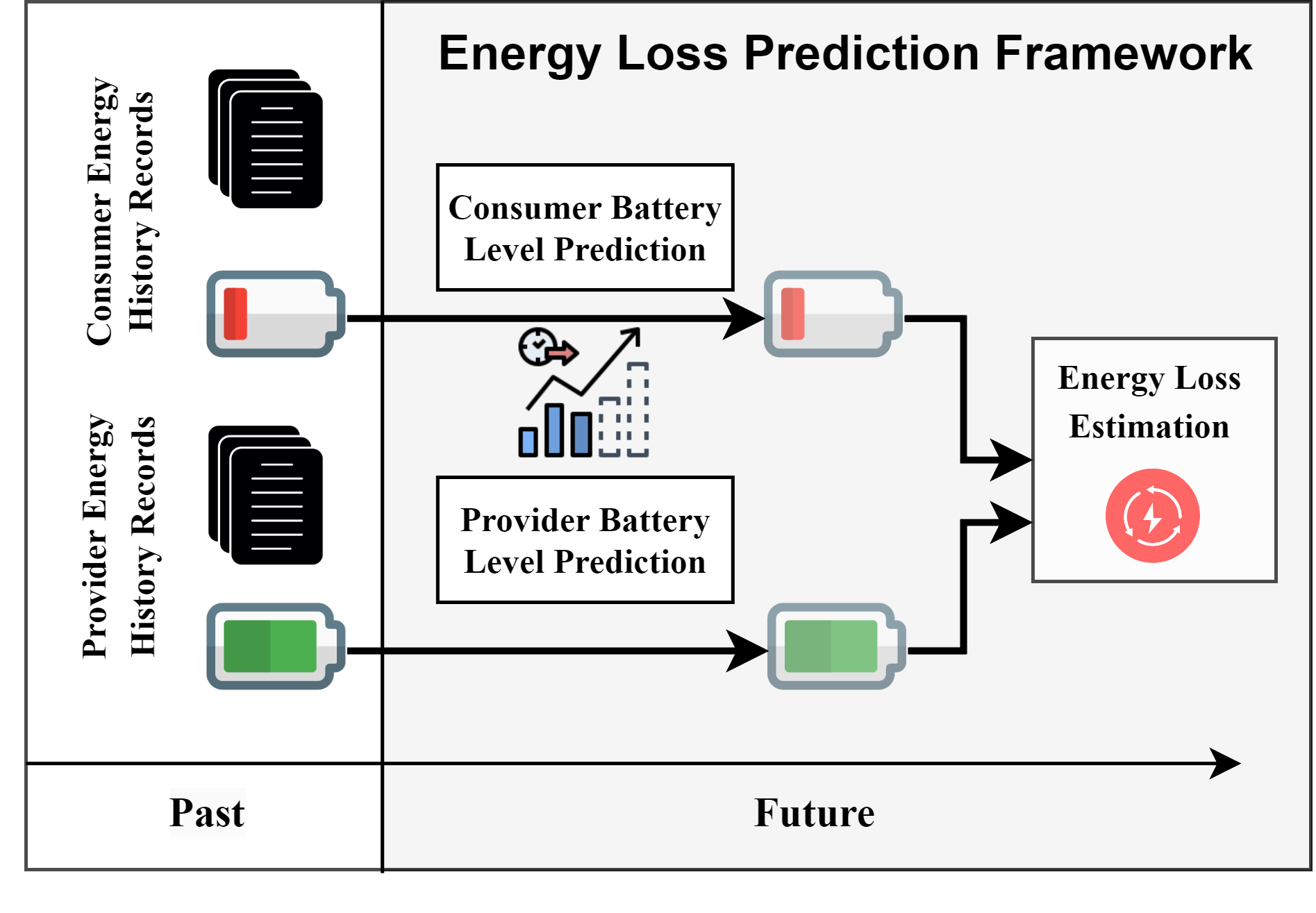}   
    \caption{High-level Energy Loss Prediction (ELP) framework}        
    \label{fig:Hframework}
\end{figure}

We propose a novel Energy Services Loss Prediction framework (\textit{ELP}) to \textit{estimate energy services transfer loss}. Our framework estimates energy loss by considering the following factors: (1) history of wireless charging data between two IoT users, including battery level history during wireless energy sharing, (2) history of IoT users' device energy usage, i.e., self-consumption, represented by battery levels in the idle state, and (3) wireless charging parameters such as distance and time. Our framework predicts the future battery levels of the provider and consumer (See Fig.\ref{fig:Hframework}). The predicted values were then used to estimate the  energy loss. To achieve this, our framework initially identifies abnormal charging behaviors such as outlier battery levels. Subsequently, it predicts the battery levels of providers and consumers under various charging states, including charging and idle. We utilized Informer, an efficient time series model \cite{zhou2021informer}, in the prediction phase. However, Informer resulted in low accuracy due to the fully zero initialization of the generative inference within its decoder. To address this issue, we extended the Informer model to \textit{Easeformer}. The Easeformer model achieves a higher prediction accuracy as it effectively captures the features of IoT users and their energy sharing preferences. These features form the prior knowledge of the partial generative inference. We employed a real-world wireless energy sharing dataset to demonstrate the effectiveness of the proposed framework. To the best of our knowledge, existing research has not considered the energy loss that occurs during wireless energy delivery \cite{abusafia2022services}\cite{lakhdari2020Vision}. This work represents one of the first attempts to quantify energy loss in the context of wireless energy sharing services. The main contributions of this paper are as follows:\looseness=-1

\begin{itemize} 
    \item A novel Energy Loss Prediction (ELP) framework to predict energy services transfer loss.  
    \item A novel prediction module (Easeformer) that effectively predicts the energy provider's loss and the energy consumer's gain.
    \item An Encoder Input Transformer (EIT) to identify the wireless charging patterns according to the distance between IoT users.
    \item A decoder with partial generative inference to  study the impact of various start token lengths.
    \item A comprehensive experiment to demonstrate that our energy loss prediction framework outperforms state-of-the-art time series prediction model using real collected energy services datasets.

\end{itemize}

  \begin{figure}[!t]
    \centering
     \setlength{\abovecaptionskip}{-2pt}
        \includegraphics[width=0.95\linewidth]{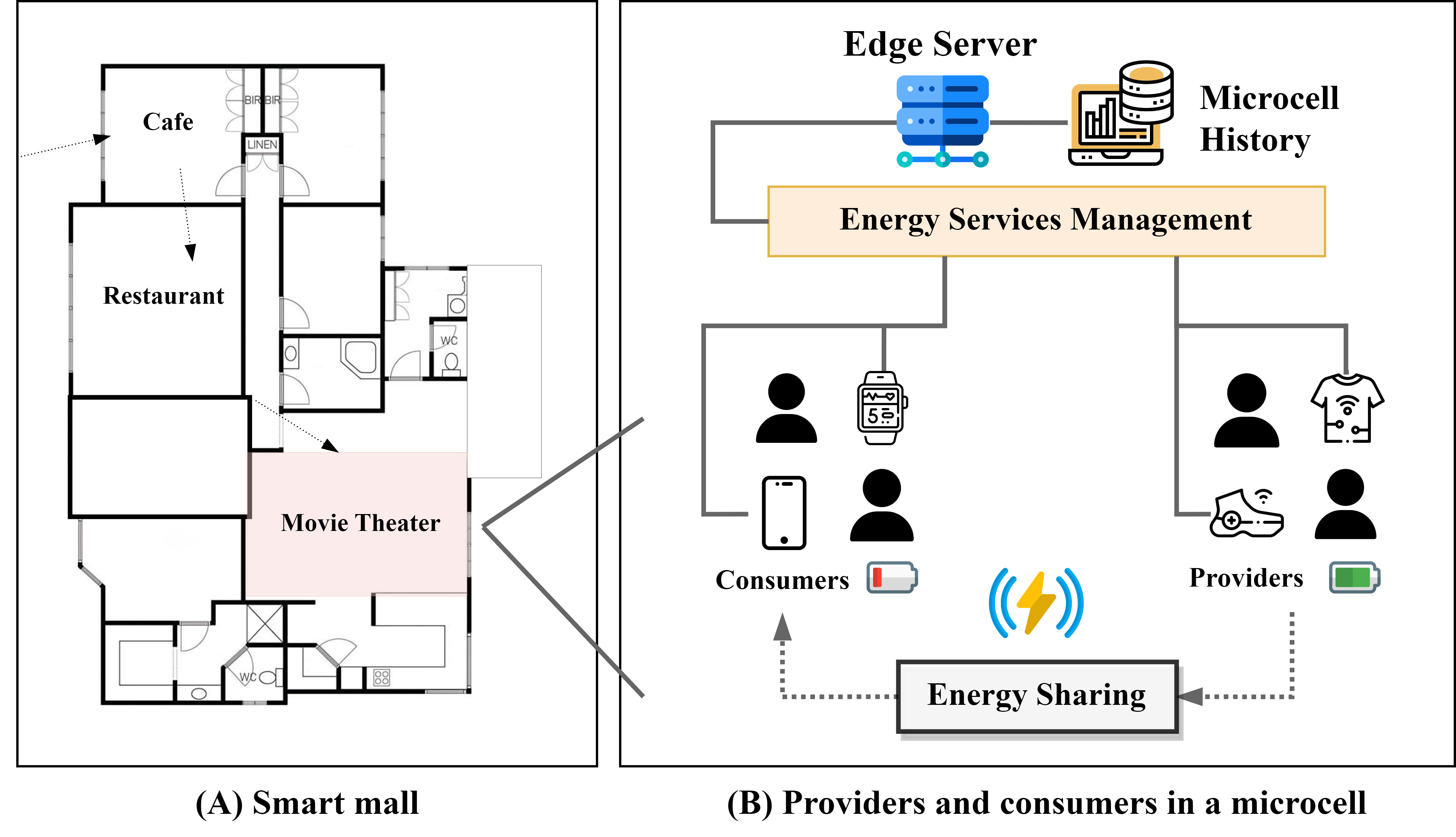}   
    \caption{The energy services environment}        
    \label{fig:scenario}\vspace{-8pt}
\end{figure}

 \begin{figure*}[!t]
    \centering
     \setlength{\abovecaptionskip}{-5pt}
    \setlength{\belowcaptionskip}{-35pt}
        \includegraphics[width=0.7\linewidth]{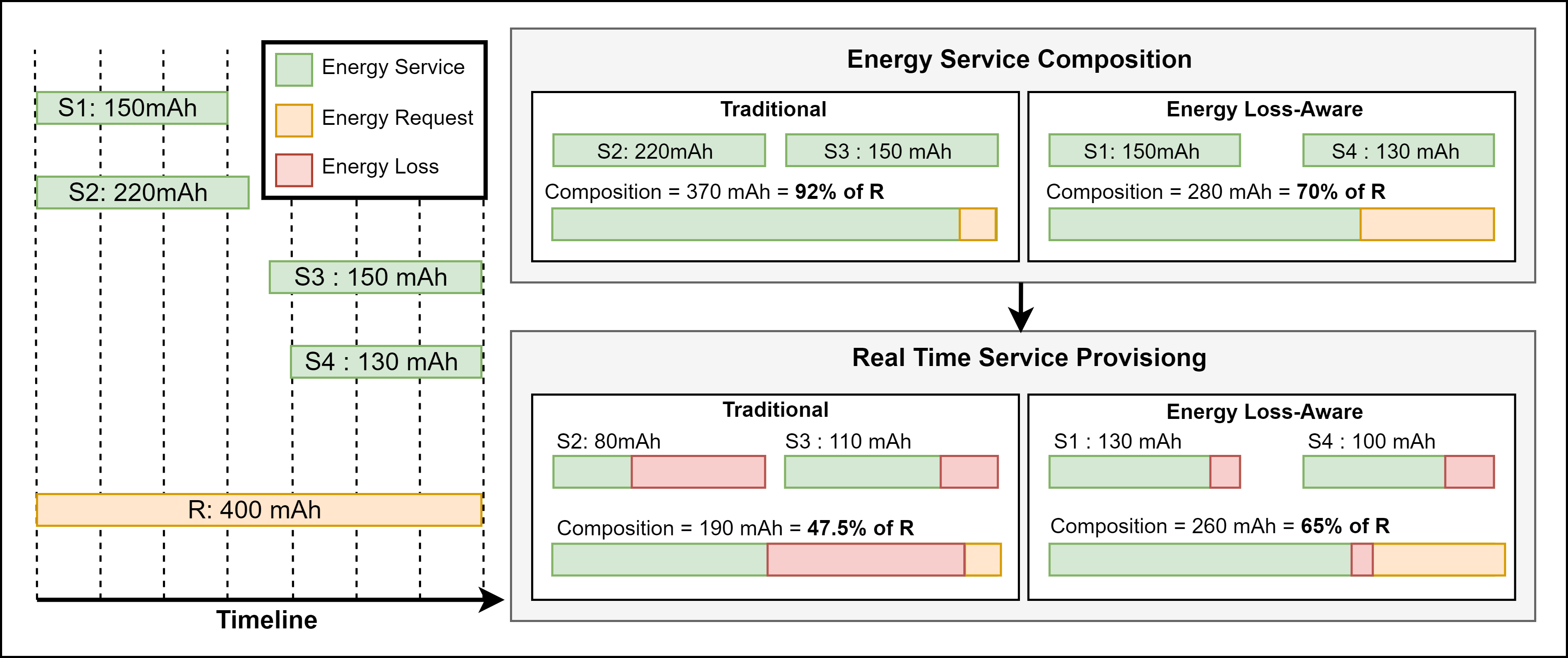}   
    \caption{Example of the motivating scenario}        
    \label{fig:Mscenario}
\end{figure*}

\vspace{-5pt}
\subsection{Motivating Scenario}

We describe a scenario in a confined area (i.e., a microcell), where people typically gather (See Fig.\ref{fig:scenario} (A)). Each microcell may have several IoT devices that act as energy consumers or providers (See Fig.\ref{fig:scenario} (B)). Consumers and providers may submit their requests or service advertisements to a local \textit{edge}, e.g., a router in a microcell. Assume that a \textit{device owner}, e.g., a \textit{smartphone}, requires energy to execute some critical tasks, such as making a call. In this scenario, the consumer is assumed to be able to receive service advertisements from multiple energy providers through the edge. We assume that consumers can receive energy from one provider at a time. Several resource-allocation frameworks have been proposed to select the optimal set of energy services for a consumer \cite{lakhdari2020composing}\cite{lakhdari2018crowdsourcing}. However, none of the existing frameworks considers the energy loss that occurs during wireless energy delivery. In reality, consumers often receive less energy than expected\cite{yang2023monitoring}\cite{yang2022towards}. In this respect, it is important to estimate the energy loss while exchanging services. Indeed, services with high energy loss do not deliver the expected amount of energy. As a result, selecting services with high energy loss will discourage consumers from participating in this ecosystem \cite{lakhdari2020Vision}.\looseness=-1

Figure \ref{fig:Mscenario} shows an example of the received energy services with different energy amounts. A traditional energy service composition allocates services with a higher capacity. For instance, using traditional composition, the first ($S1$) and third  ($S3$) services are selected with the expectation of fulfilling 92\% of the required energy. However, in real-time charging, energy loss occurs based on several factors, such as distance. The energy composition fulfills 47.5\% of the required energy due to the energy loss of both services. On the other side, informing the composition framework with the predicted energy loss will enhance the selection of services. For instance, given the energy loss of shared services, the energy-loss-aware composition will guarantee fulfilling 65\% of the required energy. In this case, the energy loss plays a key role in the development of composition algorithms \cite{lakhdari2020Vision}. Therefore, estimating the energy loss before selecting and composing energy services ensures the efficient composition of the services.\looseness=-1

We focus on estimating the energy loss based on the history data of IoT users stored on the edge. Based on these history data, we designed and developed a machine-learning-based framework to predict the energy loss. The predicted energy loss may be considered as a key indicator for selecting services in the energy-sharing process. In the future, we aim to design an energy loss-aware service composition. To the best of our knowledge, \textit{this work is among the first attempts to capture energy loss while exchanging wireless energy services}.\looseness=-1


\section{Related Work}
The background of our work comes from three areas: energy sharing services, energy consumption analysis, and machine learning-based energy consumption prediction. We present the related work to our research in the three domains.\looseness=-1

Energy sharing services have emerged as a novel approach for charging nearby IoT devices \cite{abusafia2022services}\cite{lakhdari2020Vision}. Several studies have proposed solutions to meet the requirements of energy consumers \cite{lakhdari2018crowdsourcing}\cite{lakhdari2020elastic}\cite{lakhdari2020fluid}. A temporal composition was proposed to maximize the energy provided using a fractional knapsack \cite{lakhdari2018crowdsourcing}. An elastic composition algorithm was proposed to address the highly fluctuating reliability of the energy providers \cite{lakhdari2020elastic}. This algorithm  selects more reliable services by prolonging a consumer's stay using   the concepts of soft and hard deadlines. The fluid approach leverages crowd mobility patterns to predict intermittent disconnections in energy services and then replaces or tolerates such disconnections \cite{lakhdari2020fluid}. Another study proposed the use of energy services as a tool for increasing consumer satisfaction \cite{abusafia2022maximizing}\cite{Amani2022QoE}. Other studies have addressed challenges from the provider’s perspective \cite{abusafia2020incentive}\cite{abusafia2020Reliability}. A context-aware incentive model was introduced to overcome resistance to providing energy services \cite{abusafia2020incentive}. An energy-composition framework considering consumer reliability has been suggested to encourage providers to share their energy \cite{abusafia2020Reliability}. To the best of our knowledge, none of the existing research considers the energy loss that occurs while transferring energy \cite{abusafia2022services}\cite{lakhdari2020Vision}. \textit{This work is among the first attempts to capture energy loss in the context of wireless energy sharing services}.\looseness=-1


Traditionally, energy consumption analysis has been categorized into hardware-based profiling and software-based profiling \cite{AHMAD201542}. Hardware-based profiling employs external hardware equipment such as a multimeter to measure the current and voltage, thereby estimating the power consumed by a smartphone during system activities. However, the experimental results may vary due to differences in the experimental devices and research methodologies. For example, in a study involving Google Nexus S, researchers removed the lithium-ion battery and used an external DC power source with a fixed voltage to focus on current. In this case, the screen display, GPS, and Wi-Fi were identified as the most power-consuming modules \cite{6920375}. In contrast, another study using Nokia N95 found that wireless technologies consume the most energy \cite{5956528}. Software-based profiling primarily involves measurements obtained through software programs. One such example is a mobile phone energy-monitoring system designed to analyze energy consumption at the application level \cite{6142190}. This system was developed to accurately record the energy consumption of each application and rank them accordingly.\looseness=-1


Machine learning-based prediction methods employ cutting-edge machine learning (ML) algorithms to forecast energy consumption. For example, one study utilized \textit{GreenMiner}, an Android device energy usage recorder, to gather datasets for subsequent profiling \cite{8094428}. Given that the collected data were time series, researchers developed a model using recurrent neural networks (RNNs) and implemented a long short-term memory (LSTM) algorithm. Additionally, they employed a support vector machine (SVM), shallow multi-layer perceptron (MLP), and linear regression for comparison. The results indicated that time series models, such as LSTM, were more effective in predicting the energy consumption. With the ongoing advancements in transformer-based algorithms, Informer, a state-of-the-art time series model, has been introduced \cite{zhou2021informer}. The Informer boasts superior algorithmic efficiency compared to previous time series models.\looseness=-1

\section{System Model}
\label{systemmodel}

We present definitions of energy consumers and providers. We then introduce a formal presentation of the energy loss. This study considers a provisioning framework for stationary services and requests. The goal is to accurately predict the energy loss that may occur during the energy sharing process. We use the below definitions to formulate the problem.\looseness=-1

\subsection{Preliminaries}

\begin{definition}\label{define:EHR}
\textit{\textbf{Energy History Record (EHR)}} is the history of an IoT user, i.e., consumer or provider, where the history includes the battery levels during a \textit{single} period of time. The history can be for the device while sharing energy or in daily usage, i.e., without sharing energy. The history \textit{is defined as a tuple of $<rid, uid, state, d, user\_type, BL>$  where:}
 \begin{itemize}[ noitemsep,nosep,leftmargin=8pt,labelsep=4pt,itemindent=2pt]
    \item $rid$ is a unique record identifier,
    \item $uid$ is a unique user identifier,
    \item $state$ is the energy state of the  user which can be idle, i.e., any state of usage except wireless sharing,  or sharing, i.e., during an energy sharing process,
    \item $d$ the distance between the consumer and provider in the state of wireless energy sharing,
    \item $user\_type$ is the type of user, i.e., consumer or provider in the state of wireless energy sharing,
     \item $BL$ is a set of $\{BL_{t_0}, BL_{t_1}, …, BL_{t_k}\}$ where $BL_{t_i}$ is the IoT battery level at time $t_i$, $t_{0}$ is the start time and $t_{k}$ is the end time. 
 \end{itemize}
\end{definition}

\begin{definition}\textit{\textbf{EaaS Consumer (C)}} represents the user history profile as a consumer and is formulated using $n$ energy  history records $EHR$. \textit{EaaS Consumer} \textit{is defined as a tuple of $<\mathcal{CB}, \mathcal{NCB}, \mathcal{CG}, \mathcal{CU}>$  where:}
 \begin{itemize}[ noitemsep,nosep,leftmargin=8pt,labelsep=4pt,itemindent=2pt]
     \item $\mathcal{CB}$ is a set of $\{CB^{t}_{1}, CB^{t}_{2}, …, CB^{t}_{n}\}$ where $n$ is the number of wireless energy sharing history records $EHR$ and  $CB^{t}_{i}$ is the consumer battery level at time $t$ from the history record $EHR_i$,
     \item $\mathcal{NCB}$ is a set of $\{NCB^{t}_{1}, NCB^{t}_{2}, …, NCB^{t}_{n}\}$ where $NCB^{t}_{i}$ is the consumer battery level in idle state (not charging) at time $t$ from the history record $EHR_i$,
     \item $\mathcal{CG}$ is a set of $\{CG^{t}_{1}, CG^{t}_{2}, …, CG^{t}_{n}\}$ where $CG^{t}_{i}$ is the  consumer gain from the energy sharing process at time $t$ from the history record $EHR_i$, $CG^{t}_{i}$ is computed using the consumer battery level in sharing state $CB\in \mathcal{CB}$ as follows:\looseness=-1
     \begin{equation}
     \label{eq:CG}
        CG^{t}_{i} = CB^{t}_{i} - CB^{t_0}_{i}
     \end{equation}
     \item $\mathcal{CU}$ is a set of $\{CU^{t}_{1}, CU^{t}_{2}, …, CU^{t}_{n}\}$ where $CU^{t}_{i}$ is the consumer usage, i.e., self-consumption, at time  $t$ from the history record $EHR_i$, $CU^{t}_{i}$ is computed using the consumer battery level in idle state $NCB\in \mathcal{NCB}$ as follows:
     \begin{equation}
     \label{eq:CU}
        CU^{t}_{i} = NCB^{t_0}_{i} - NCB^{t}_{i}
     \end{equation}
 \end{itemize}
\end{definition}

\begin{definition}\textit{\textbf{EaaS Provider (P)}} represents the user history profile as a provider and is formulated using $n$ energy  history records \textit{EHR}. \textit{EaaS Provider} \textit{is defined as a tuple of $<\mathcal{PB}, \mathcal{NPB}, \mathcal{PL}, \mathcal{PU}>$  where:} \begin{itemize}[ noitemsep,nosep,leftmargin=8pt,labelsep=4pt,itemindent=2pt]
     \item $\mathcal{PB}$ is a set of $\{PB^{t}_{1}, PB^{t}_{2}, …, PB^{t}_{n}\}$ where  $n$ is the number of energy sharing history records (EHR) and $PB^{t}_{i}$ is the provider battery level at time $t$ from  the history record $EHR_i$,
     \item $\mathcal{NPB}$ is a set of  $\{NPB^{t}_{1}, NPB^{t}_{2}, …, NPB^{t}_{n}\}$ where $NPB^{t}_{i}$ is the  provider battery level in idle state (not charging) at time $t$ from  history record $EHR_i$,
     \item $\mathcal{PL}$ is a set of  $\{PL^{t}_{1}, PL^{t}_{2}, …, PL^{t}_{n}\}$ where $PL^{t}_{i}$ is the provider loss from the energy sharing process at time $t$ from  history record $EHR_i$, $PL^{t}_{i}$ is computed using the provider battery level in sharing state $PB\in\mathcal{PB}$ as follows:
     \begin{equation}
     \label{eq:PL}
        PL^{t}_{i} = PB^{t_0}_{i} - PB^{t}_{i}
     \end{equation}
     \item $\mathcal{PU}$ is a set of $\{PU^{t}_{1}, PU^{t}_{2}, …, PU^{t}_{n}\}$ where $PU^{t}_{i}$  is the provider usage, i.e., self-consumption, at time $t$ from  history record $EHR_i$, $PU^{t}_{i}$ is computed using the provider battery level in idle state $NPB\in\mathcal{NPB}$ as follows:
     \begin{equation}
     \label{eq:PU}
        PU^{t}_{i} = NPB^{t_0}_{i} - NPB^{t}_{i}
     \end{equation}
 \end{itemize}
\end{definition}

\begin{definition}\label{definition}\textit{\textbf{Energy Loss ($\mathcal{EL}$)}} \textit{is a set of  $\{EL^{t}_{1}, EL^{t}_{2}, …, EL^{t}_{n}\}$, $EL^{t}_{i}$ is the energy loss of the wireless transfer without the energy consumed by the device for other purposes. $EL^{t}_{i}$ is computed as follows:}
\begin{equation}
 \label{eq:el}
 EL^{t}_{i}  = RT^{t}_{i} - RR^{t}_{i}
  \vspace{-5pt}
\end{equation}
 \begin{itemize}[ noitemsep,nosep,leftmargin=8pt,labelsep=4pt,itemindent=2pt]
     \item $RT^{t}_{i}$ represents the actual transferred energy service at time $t$ from the energy history record $EHR_i$. $RT^{t}_{i}$ is computed using the provider loss $PL\in \mathcal{PL}$ and usage $PU\in \mathcal{PU}$ as follows:
     \begin{equation}
     \label{eq:RT}
     \begin{aligned}
     RT^{t}_{i}  = PL^{t}_{i} - PU^{t}_{i}
      \end{aligned}
     \end{equation}

     \item $RR^{t}_{i}$ represents  the actual received energy service at time $t$ from the energy history record $EHR_i$. $RR^{t}_{i}$ is computed using the consumer gain $CG\in \mathcal{CG}$ and usage $CU\in \mathcal{CU}$ as follows:
     \begin{equation}
     \label{eq:RR}
     \begin{aligned}
     RR^{t}_{i}  = CG^{t}_{i} + CU^{t}_{i}
      \vspace{-5pt}
      \end{aligned}
     \end{equation}
 \end{itemize}
\end{definition}


\subsection{Problem formulation}
Given the history of two IoT users, i.e., a consumer and a provider, where each user history record $EHR$ consists of $m$ energy \textit{usage} history records and $n$ energy \textit{sharing} history records. Each history record $EHR$ includes the battery levels of the IoT user over a period of time $T^t_i$. $T^t_i$ is a time series of $<T^{t_0}_i, T^{t_1}_i, …, T^{t_k}_i>$  where $T^{t_j}_i$ is a timestamp $j$ in the energy history record $EHR_i$. A set of $T_i$ creates  $\mathcal{T}$, i.e., a  set of $n$ time series. $\mathcal{T}$ is defined based on the user $n$ number of $EHR$. Therefore, $\mathcal{T}$ is a set of $ \{T^t_1, T^t_2, …, T^t_{n}\}$. For simplicity, we use  $t_0$ instead of $T^{t_0}_i$ and $t_{k}$ instead of $T^{t_{k}}_i$ in the following definitions. Moreover, given the energy sharing distance $\mathcal{D}$ as a set of $\{d_1, d_2, …, d_{n}\}$ where $d_i$ is the energy sharing distance between a provider and a consumer from $EHR_i$. We transform the problem of energy loss estimation into the integration of four parallel time series forecasting problems\cite{zhou2021informer}\cite{cho2014properties}\cite{sutskever2014sequence}. The goal of the four aforementioned problems is to predict provider loss $\mathcal{PL}$, provider usage $\mathcal{PU}$,  consumer gain $\mathcal{CG}$, and consumer usage $\mathcal{CU}$. Each predicted  time series will be used later to compute the energy loss. Each time series forecasting problem can be formulated as follows:\looseness=-1
\begin{itemize}
    \item Given the input $\mathcal{X}^t = \{x^t_1,...,x^t_{L_x}\mid x^t_i\in\mathbb{R}^{d_x}\}$ at time $t$ where $L_{x}$ is the  length of the input, and $\mathcal{X}^t$ is the flattened  sequence from the $n$ number of $EHR$.
    \item The objective is to predict the  corresponding output sequence $\mathcal{Y}^t = \{y^t_1,...,y^t_{L_y}\mid y^t_i\in\mathbb{R}^{d_y}\}$ where $L_{y}$ is the length of the outupt.
\end{itemize}

 \begin{figure*}[!t]
    \centering
        \includegraphics[width=0.82\linewidth]{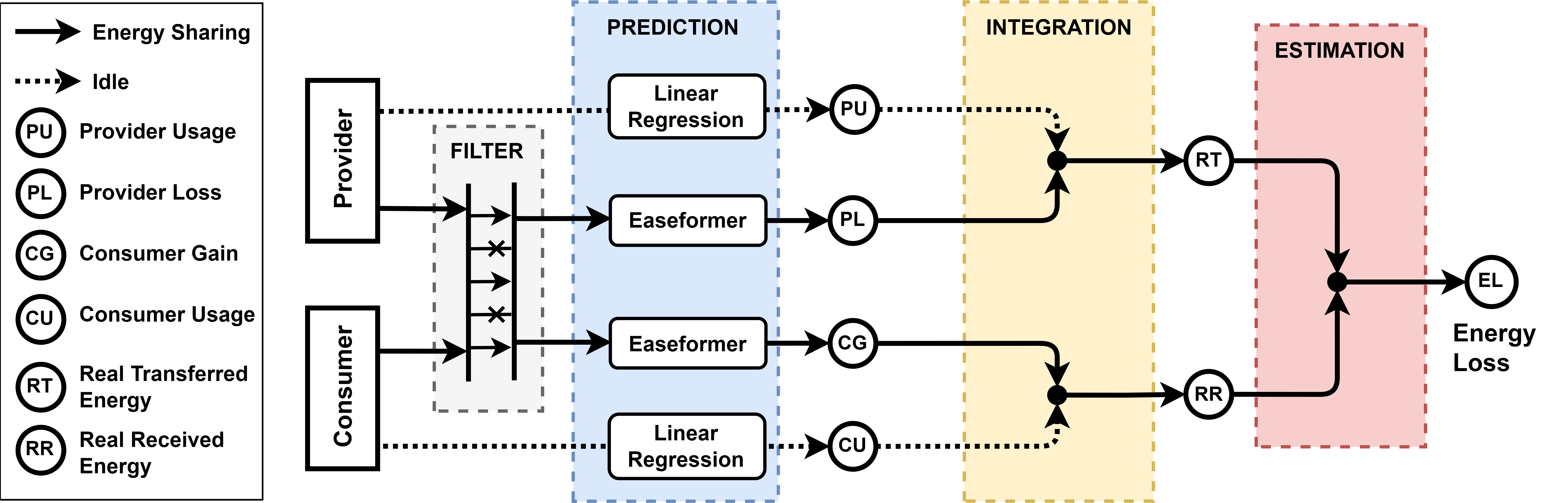}
   
    \caption{Energy Loss Prediction (ELP) framework overview}
        
    \label{fig:systemmodel}
\end{figure*}
\noindent We use the following assumptions to formulate the problem.
\begin{itemize}
\item Consumers and providers  are committed to the energy sharing process, i.e., the history records are for completed energy services delivery.
\item Providers are motivated  to share their energy using an incentive model \cite{abusafia2022services}\cite{abusafia2020incentive}.
\item Providers and consumers are static during energy sharing.
\item The energy sharing distance between the provider and consumer is prior knowledge, i.e., the energy sharing distance is known in advance.
\item The self-consumption of providers and consumers is stable and may not change a lot in both conditions of charging or idle \cite{carroll2010analysis}.
\item IoT users prefer longer energy sharing distances.
\item A secure and trustworthy framework is utilized to preserve the privacy and security of IoT devices\cite{ba2022multi}.
\end{itemize}

\vspace{-1pt}
\section{Energy Loss Prediction Framework}
\label{section}


We present an Energy Loss Prediction $ELP$ framework to predict the energy loss that occurs in sharing energy services. As previously mentioned, our framework estimates energy loss by predicting the future battery levels of the provider and consumer. The predicted values were then utilized to estimate the energy loss (See Fig.\ref{fig:Hframework}). In detail, the framework consists of four phases (See Fig.\ref{fig:systemmodel}): (1) Filter, (2) Prediction, (3) Integration, and (4) Estimation. In the first phase, the history energy sharing data are filtered, and abnormal battery-level data are removed. The second phase uses the Easeformer and linear regression models to predict  provider usage, provider loss, consumer usage, and consumer gain. The third phase integrates the predicted values from the previous phase to compute the actual  transferred and received energy. The final phase uses the computed actual transferred and received energy to estimate the energy loss. Algorithm \ref{alg:two} presents the four phases for implementing the ELP framework. In what follows, we present each phase in detail.  


\subsection{ELP Filter Phase}

This phase involves the following two steps: (1) compute the consumer gain and the provider loss from the raw history data of the provider and consumer (2) remove the abnormal data, i.e., outliers, from the computed consumer gain and provider loss. Typically,  outliers are defined as values that deviate significantly from the majority of the battery level data points \cite{ester1996density}. In the first step, given energy sharing raw history data $\mathcal{\Tilde{CB}}$ and $\mathcal{\Tilde{PB}}$, we use Eq.\ref{eq:CG} to compute the consumer gain $\mathcal{\Tilde{CG}}$ and Eq.\ref{eq:PL} to compute the provider loss $\mathcal{\Tilde{PL}}$ (See Algorithm \ref{alg:two}, Line 1). In the second step, we filter the computed consumer gain $\mathcal{\Tilde{CG}}$ and provider loss $\mathcal{\Tilde{PL}}$. The reason for filtering is that, in the prediction phase,  we used the Mean Squared Error (MSE) as the loss function. It is important to note that $MSE$ is vulnerable to outliers \cite{li2018predicting}. As a result, we utilize an outlier detection technique called Density-Based Spatial Clustering of Applications with Noise (DBSCAN) as a filter to eliminate  outliers \cite{ester1996density}. Upon examining our collected dataset, we observed that the likelihood of errors in the wireless energy sharing process  is low,  leading to rare occurrences of abnormal battery-level data. Consequently, these anomalous data points exhibit a lower density than normal data points. Unlike statistical methods that detect anomalous points above or below a specific threshold, i.e., extremes, DBSCAN effectively identifies infrequently occurring data i.e., detecting outlier points with lower density \cite{ccelik2011anomaly}. Therefore, we employ DBSCAN to detect and remove outliers. The inputs for the filter phase are the raw energy sharing history data of the provider $P$ and consumer $C$. The outputs of this phase are  the cleaned provider loss $\mathcal{PL}$ and consumer gain $\mathcal{CG}$ (See Algorithm \ref{alg:two}, Lines 1-2).\looseness=-1

\begin{algorithm}[!t]
\caption{Energy Loss Prediction Framework (ELP)}\label{alg:two}



\textbf{Input:} $P$: provider history profile, $C$: consumer history profile\\
\textbf{Output:} $\mathcal{\hat{EL}}$: predicted energy loss

    \begin{algorithmic}[1]
    \Statex \textbf{Phase 1: Filter}
    \STATE $\mathcal{\Tilde{PL}},\mathcal{\Tilde{CG}}\leftarrow \mathbf{Compute\_Energy}(P.\mathcal{\Tilde{PB}},C.\mathcal{\Tilde{CB}})$
    \STATE $\mathcal{PL},\mathcal{CG}\leftarrow \mathbf{DBSCAN}(\mathcal{\Tilde{PL}},\mathcal{\Tilde{CG}})$
    \Statex \textbf{Phase 2: Prediction}
    \STATE $\mathcal{\hat{PL}},\mathcal{\hat{CG}}\leftarrow \mathbf{Easeformer}(\mathcal{PL},\mathcal{CG})$
    \STATE $\mathcal{\hat{PU}},\mathcal{\hat{CU}}\leftarrow \mathbf{LinearRegression}(\mathcal{PU},\mathcal{CU})$
    \Statex \textbf{Phase 3: Integration}
    \STATE $\mathcal{\hat{RT}},\mathcal{\hat{RR}}\leftarrow \text{Integration of } \mathcal{\hat{PL}}, \mathcal{\hat{PU}},\mathcal{\hat{CG}}, \mathcal{\hat{CU}}$
    \Statex \textbf{Phase 4: Estimation}
    \STATE $\mathcal{\hat{EL}}\leftarrow \text{Energy loss estimation based on } \mathcal{\hat{RT}},\mathcal{\hat{RR}}$
    \STATE \textbf{return } $\mathcal{\hat{EL}}$
    \end{algorithmic}
\end{algorithm}
\setlength{\textfloatsep}{5pt}

\subsection{ELP Prediction Phase}
This phase aims to predict the future consumer gain, consumer usage, provider loss, and provider gain using the calculated values of these attributes from the previous phase. As previously mentioned, the predicted values will be used in the following phase to compute energy loss. In this phase, we employ two types of models, i.e., Transformer-based time series model and linear regression, due to the distinct nature of data between the energy sharing and idle states. Consequently, we utilize a multi-head attention-based time series model, Informer, as our baseline to capture the consumer gain and the provider loss values separately in the energy sharing state \cite{lu2015wireless}. The Informer algorithm employs ProbSparse multi-head self-attention, which reduces the time and space complexity to $\mathcal{O}(L\log{}L)$. This is significantly lower than the standard Transformer's $\mathcal{O}(L^2)$ for each layer, where $L$ represents the length of inputs/outputs. As a result, Informer is more suitable for handling time series forecasting problems \cite{zhou2021informer}. However, the Informer delivered low accuracy due to the fully zero initialization of the generative inference within its decoder. To address this issue, we extend Informer to \textit{Easeformer}, which achieves higher prediction accuracy by effectively capturing the features of IoT users and their energy sharing preferences. Thus, Easeformer is used to predict the values of provider loss and consumer gain in the energy sharing state. On the other hand, we employ linear regression for predicting consumer and provider usage in the idle state (See dotted lines in Fig.\ref{fig:systemmodel}) \cite{maulud2020review}. This is based on our assumption that battery levels in the idle state, i.e., self-consumption, remain stable and do not change significantly. We formulated this assumption in light of the battery level variation trend discussed in \cite{carroll2010analysis}. In summary, the prediction phase utilizes two Easeformer models and two linear regression models to predict energy values changes of providers and consumers in each state, i.e., energy sharing and idle states, respectively. In the following subsections, we present each model in detail.\looseness=-1


 \begin{figure}[!t]
    \centering
     \setlength{\abovecaptionskip}{-2pt}
    \setlength{\belowcaptionskip}{-30pt}
        \includegraphics[width=0.97\linewidth]{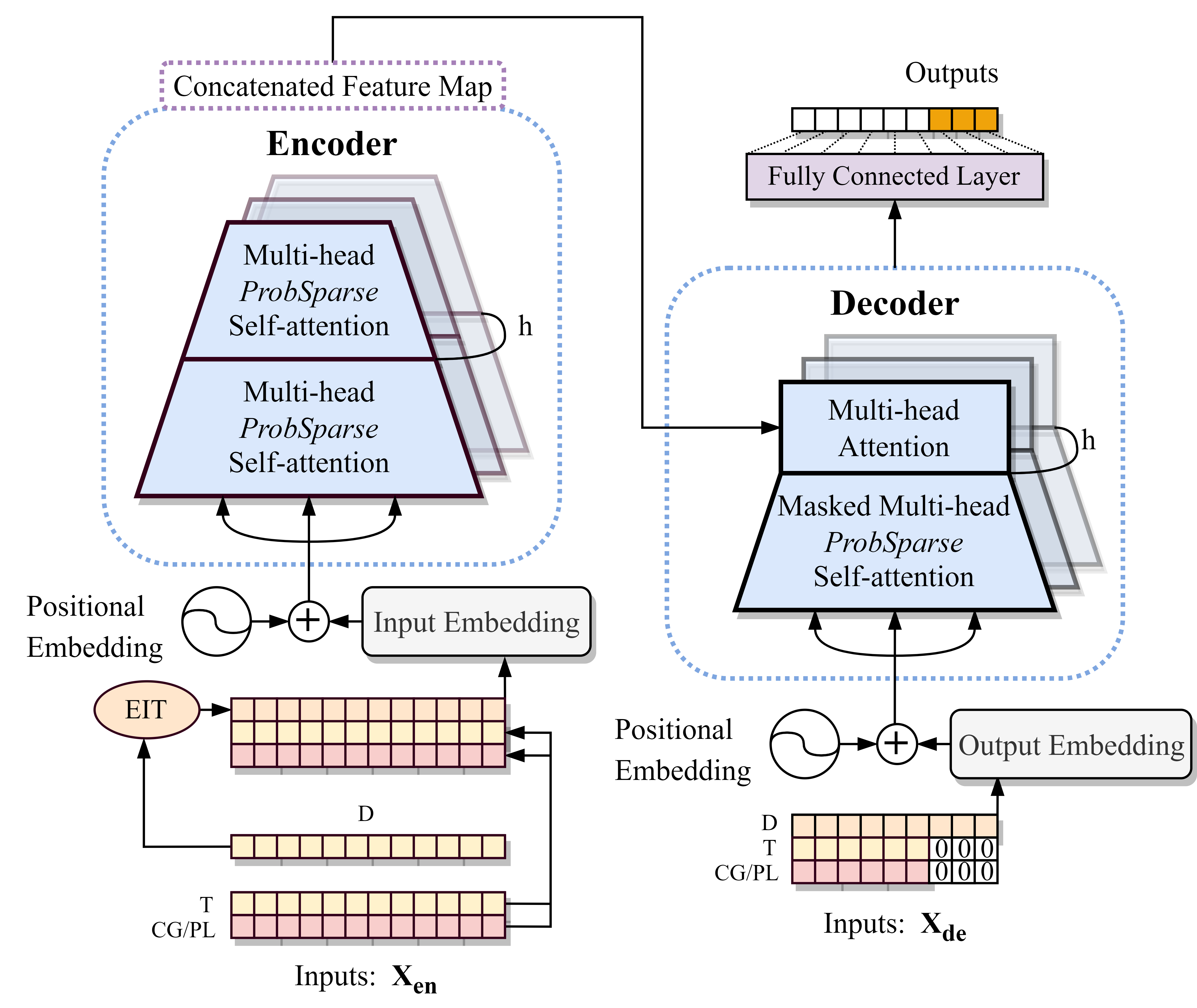}

\caption{Overview of the Easeformer model. On the left: The encoder processes an extended, long sequence of multidimensional inputs, where each input dimension is represented by unique colors (for example, the orange series signifies the energy sharing distance). The blue trapezoid represents a self-attention distilling operation that extracts dominant attention while reducing the network size. On the right: The decoder accepts a long-sequence input, padding target elements to zero for dimensions lacking prior knowledge. In the case of dimensions with prior knowledge, such as energy sharing distance, the decoder makes use of this prior information. The decoder subsequently computes the weighted attention composition of the feature map and generates output elements through a generative process \cite{zhou2021informer}.}

        
    \label{Easeformer}
\end{figure}

\subsubsection{Easeformer model for sharing state prediction}

The Easeformer model overview is depicted in Fig.\ref{Easeformer}. We consider the consumer gain $\mathcal{CG}$ and the provider loss $\mathcal{PL}$ as the model targets.  We then use the energy sharing distance $\mathcal{D}$  and the energy sharing time $\mathcal{T}$ as input features. Therefore, $\mathcal{CG}$/$\mathcal{PL}$, $\mathcal{D}$, and $\mathcal{T}$ are concatenated as $\bold{X}^t_{en} = \{x^t_1,...,x^t_{L_{x}}\mid x^t_i\in\mathbb{R}^{L_{x}\times d_{model}}\}$, i.e., the input data of the encoder (See Fig.\ref{Easeformer}).\looseness=-1

\subsubsection*{\textbf{Easeformer Encoder}}

In the encoder structure of the Easeformer, we assume that IoT users prefer longer energy sharing distances. Moreover, based on the human decision model used in \cite{reverdy2015parameter},  we propose an Encoder Input Transformer (\textit{EIT}). EIT applies the softmax function to the feature $\mathcal{D}$ to obtain the probabilities $\mathbf{P}_{\text{user}}$ for various distances that IoT users are likely to select (See Fig.\ref{Easeformer}). We then normalize the distance set to a new set ranging from 0 to 1. This allows IoT users to choose the longest energy sharing distance with higher probability. Additionally, we employ the softmax temperature $\tau$ to fine-tune the smoothness of the output probability distribution \cite{hinton2015distilling}. When $\tau$ $\rightarrow$ $1$, the function becomes equivalent to a traditional softmax function. When $\tau$ $\rightarrow$ $0$, the output distribution converges to a mass point. When $\tau$ $\rightarrow$ $\infty$, all the elements in $\mathbf{P}_{\text{user}}$ become equal, resulting in a smooth approximated distribution. Consequently, the probability of an IoT user selecting the $i$-th distance among $k$ different distances can be computed as follows: 
\vspace{-5pt}

     \begin{equation}
        \mathbf{{P}}_{user}(d_i\mid d_1...d_{k}) = \frac{\mathbf{exp}(d_i/\tau)}{\sum_{j=1}^{k}\mathbf{exp}(d_j/\tau)} 
    \vspace{-5pt}
     \end{equation}
Algorithm \ref{alg:one} presents the aforementioned process in detail.

Moreover, in this paper, we do not focus on the local temporal context of time series inputs. Therefore, we remove the time representation from the input representation part compared to the Informer embedding structure and retain the value and positional representation (See Fig.\ref{Easeformer}). After embedding, all the input data are projected into 512 dimensions and then concatenated. Consequently, the input representations $\bold{X}_{en}$ are encoded into the hidden state representations $H^t = {h^t_1,...,h^t_{L_{x}}}$. Subsequently, $H^t$ serves as an input for the ProbaSparse multi-head attention mechanism, which is defined as:\looseness=-1
\begin{algorithm}[!t]
\caption{Enocder Input Transformer (EIT)}\label{alg:one}

\textbf{Input:} $\mathcal{D}$: Energy sharing distance\\
\textbf{Output:} $\mathbf{P}_{user}$: user energy sharing distance preference probability, $dict$: a dictionary that concatenated by $\mathbf{P}_{user}$ and unique energy sharing distance $\mathcal{D}_{unique}$

    \begin{algorithmic}[1]
    \STATE $\mathbf{P}_{user} \leftarrow \phi$
    \STATE $\mathcal{D}_{unique}\leftarrow \text{select unique values in input feature } \mathcal{D}$
    \STATE $\mathbf{P}_{user} \leftarrow \mathbf{Softmax}(\mathcal{D}_{unique})$ with temperature
    \STATE $dict \leftarrow \text{concatenate $\mathcal{D}_{unique}$, $\mathbf{P}_{user}$ to a dictionary}$  
    \STATE \textbf{return } $\mathbf{P}_{user}$, $dict$
    \end{algorithmic}
\end{algorithm}

     \begin{equation}
     \begin{aligned}
        head_i = \mathbf{Softmax}\left(\frac{(h^t)^T\overline{W}^Q_i(h^t)^TW^{K^T}_i}{\sqrt{d}}\right)(h^t)^TW^V_i\\
        \mathcal{A}(h^t) = \mathbf{Concat}(head_1,...,head_h)
    \end{aligned}
    \vspace{-5pt}
     \end{equation}
where $W^Q_i$, $W^K_i$, and $W^V_i \in\mathbb{R}^{d \times d}$ are three projection matrices. The multi-head attention for $h^t$ is denoted by $\mathcal{A}(h^t)$, where  $h$ is the number of heads. $\overline{W}^Q_i$ represents the sparse matrix that contains only the Top-$u$ queries under the sparsity measurement $M(\bold{q}_i, \bold{K})$. The aforementioned query sparsity measurement is called the max-mean measurement, where $\bold{q}_i$ denotes the $i$-th row in $\bold{Q}$ \cite{zhou2021informer}. Furthermore, we use the self-attention distilling technique  to handle the redundancy of the value $\bold{V}$ in the encoder's feature map. The technique aims to privilege the superior ones with dominant features and create a focused self-attention feature map in the next layer \cite{zhou2021informer}\cite{yu2017dilated}.\looseness=-1


\subsubsection*{\textbf{Easeformer Decoder}}
In the decoder structure of the Easeformer, we employ a standard decoder structure (See Fig.\ref{Easeformer}), which is composed of a stack of two identical multi-head attention layers \cite{vaswani2017attention}. Unlike the Informer which fully  initializes the target sequence by zero, we propose a partial generative inference that makes use of prior knowledge, such as the energy sharing distance. Note, the energy distance can be computed using the technique described in \cite{jia2022smartphone}. Specifically, the partial generative inference samples an $L_{token}$-long sequence in the input sequence for features without prior knowledge. This sequence then serves as a clue to the inference of the $L_y$-length prediction. Therefore, we use the prior knowledge which is the energy distance in our context, as a partial input of the $L_y$ prediction, and we initialize other features to 0. For example, when predicting 30 points, i.e., the prediction of a 30-minute energy sharing process, we will use the known two 30-minute energy sharing history data, i.e., 60 points, and the prior knowledge, i.e., the energy sharing distance, as the start token. We then feed the generative-style inference decoder with the following vectors as follows:\looseness=-1
\begin{equation}
     \begin{aligned}
\bold{X}^t_{de_{partial}} = \mathbf{Concat}(\bold{X}^t_{token_{partial}},\bold{X}^t_0)
    \vspace{-5pt}
         \end{aligned}
     \end{equation}
\begin{equation}
     \begin{aligned}
\bold{X}^t_{de_{prior}} = \mathbf{Concat}(\bold{X}^t_{token_{prior}},\bold{X}^t_{prior})
    \vspace{-5pt}
    \end{aligned}
     \end{equation}
\begin{equation}
     \begin{aligned}
\bold{X}^t_{de} = \mathbf{Concat}(\bold{X}^t_{de_{partial}},\bold{X}^t_{de_{prior}}) 
    \vspace{-5pt}
    \end{aligned}
     \end{equation}
where $\bold{X}^t_0$ $\in$ $\mathbb{R}^{L_y\times (d_{model}-d_{prior})}$ represents the zero-initialized input tokens of the decoder, $\bold{X}^t_{prior}$ $\in$ $\mathbb{R}^{L_y\times d_{prior}}$ represents the prior knowledge-initialized input tokens of the decoder, $\bold{X}^t_{token_{partial}}$ $\in$ $\mathbb{R}^{L_{token}\times (d_{model}-d_{prior})}$ and $\bold{X}^t_{token_{prior}}$ $\in$ $\mathbb{R}^{L_{token}\times d_{prior}}$ denotes the history data that serves as a clue. Furthermore, $\bold{X}^t_{de_{partial}}$ $\in$ $\mathbb{R}^{(L_{token}+L_y)\times (d_{model}-d_{prior})}$ corresponds to features that are not known in advance, while $\bold{X}^t_{de_{prior}}$ $\in$ $\mathbb{R}^{(L_{token}+L_y)\times d_{prior}}$ represents features with prior knowledge. In the context of energy sharing, $\bold{X}^t_{de_{partial}}$ comprises the features $\mathcal{CG}$/$\mathcal{PL}$ and $\mathcal{T}$. Feature $\mathcal{D}$ is assumed to be a fixed value and known in advance, i.e., prior knowledge. Given the transformed value, i.e., $\mathbf{P}_{user}$, obtained via Algorithm \ref{alg:one}, we employ a $dict$ function to transform the prior and subsequently combine it with $\bold{X}^t_{de_{partial}}$. This results in the decoder input $\bold{X}^t_{de}$ $\in$ $\mathbb{R}^{(L_{token}+L_y)\times d_{model}}$.\looseness=-1

\subsubsection*{\textbf{Loss Function}}

With regard to the loss function of Easeformer, we choose the mean squared error loss function for the prediction, and the loss is propagated back from the decoder’s outputs across the entire model. Then, our two Easeformer models respectively output two time series predictions, i.e., $\mathcal{\hat{PL}} = \{PL^{t}_{1}, PL^{t}_{2}, …, PL^{t}_{L_{y}}\}$ and $\mathcal{\hat{CG}} = \{CG^{t}_{1}, CG^{t}_{2}, …, CG^{t}_{L_{y}}\}$ (See Algorithm \ref{alg:two}, Line 3).\looseness=-1

\subsubsection{Linear regression model for idle state prediction}

Given battery level history data in the idle state, i.e., $\mathcal{NPB}$ and $\mathcal{NCB}$. We use Eq.\ref{eq:CU} to compute the consumer usage $\mathcal{CU}$, and Eq.\ref{eq:PU} to compute the provider usage $\mathcal{PU}$. We use $\mathcal{CU}$/$\mathcal{PU}$ as the model's target. The input of the linear regression models includes the aforementioned features and time $\mathcal{T}$. We use two linear regression models,  one to predict the time series set  $\mathcal{\hat{PU}} = \{PU^{t}_{1}, PU^{t}_{2}, …, PU^{t}_{L_{y}}\}$ and another to predict the time series set $\mathcal{\hat{CU}} = \{CU^{t}_{1}, CU^{t}_{2}, …, CU^{t}_{L_{y}}\}$ (See Algorithm \ref{alg:two}, Line 4).\looseness=-1

\subsection{ELP Integration Phase}
This phase computes the real received and transferred energy using the predicted consumer gain $\mathcal{\hat{CG}}$, provider loss $\mathcal{\hat{PL}}$, consumer usage $\mathcal{\hat{CU}}$, and provider usage $\mathcal{\hat{PU}}$. The real transferred energy by the provider, denoted as  $\mathcal{\hat{RT}}$ is computed using Eq.\ref{eq:RT}, while  the real received energy by  the consumer, represented by $\mathcal{\hat{RR}}$ is computed using Eq.\ref{eq:RR}.  Overall, this phase computes $\mathcal{\hat{RT}}$ and $\mathcal{\hat{RR}}$ as shown in Algorithm \ref{alg:two} (Line 5).\looseness=-1

\subsection{ELP Estimation Phase}

The ELP Estimation Phase provides the output of the entire ELP framework. Specifically, the inputs to this phase are $\mathcal{\hat{RT}}$ and $\mathcal{\hat{RR}}$. The predicted energy loss, $\mathcal{\hat{EL}}$, is computed using Eq.\ref{eq:el}, as demonstrated in Algorithm \ref{alg:two} (Lines 6-7).\looseness=-1

\section{Experiments and Results}
  \label{ExpSection}

In this section, we first present the conducted experiment and our energy sharing datasets. We then evaluate the effectiveness of the framework in terms of the Mean Squared Error (MSE) and Mean Absolute Error (MAE). Specifically, we examine the MSE and MAE of the Easeformer and linear regression models. Furthermore, we assess the MSE and MAE of the entire ELP framework.\looseness=-1



\subsection{Dataset Description}

We collected a set of real-world wireless energy transfer datasets to train our machine learning models. Data collection was conducted using the mobile application developed by \cite{yang2022towards}. The application enables consumers to connect to a provider via Bluetooth and requests energy based on size or duration. Additionally, the app monitors the energy-sharing process between a consumer and provider by recording their battery levels at specific time intervals ($mt$), such as every 5 seconds. The granularity of the monitoring time interval $mt$ is determined by the consumer. A fine-grained time interval yields more detailed records of both consumer and provider battery levels. Furthermore, a timestamp is employed to synchronize the monitoring and recording of the energy transfer between the provider and consumer. The current version of the energy-sharing application supports a one-to-one energy transfer mode, meaning that a \textit{single} energy provider can deliver energy to only a \textit{single} energy consumer.

In our experiments, we used a Google Pixel 5 smartphone as the provider and a Google Pixel 3 as the consumer. The experiments were conducted based on the design of \cite{yang2023monitoring}. Both the consumer and provider were connected to wireless charging coils. To ensure accurate results, all experiments were carried out in a laboratory setting at an approximate temperature of 25°C. We also used the aforementioned application to request energy by time, setting the charging duration to 30 minutes and the monitoring interval $mt$ to 1 minute. For the usage (self-consumption) data collection experiments, only Bluetooth and Wi-Fi functions were activated, and the brightness level for both devices was set to the lowest setting.\looseness=-1

\begin{figure}[!t]
    \centering
     \setlength{\abovecaptionskip}{-2pt}
    \setlength{\belowcaptionskip}{-25pt}
        \includegraphics[width=0.86\linewidth]{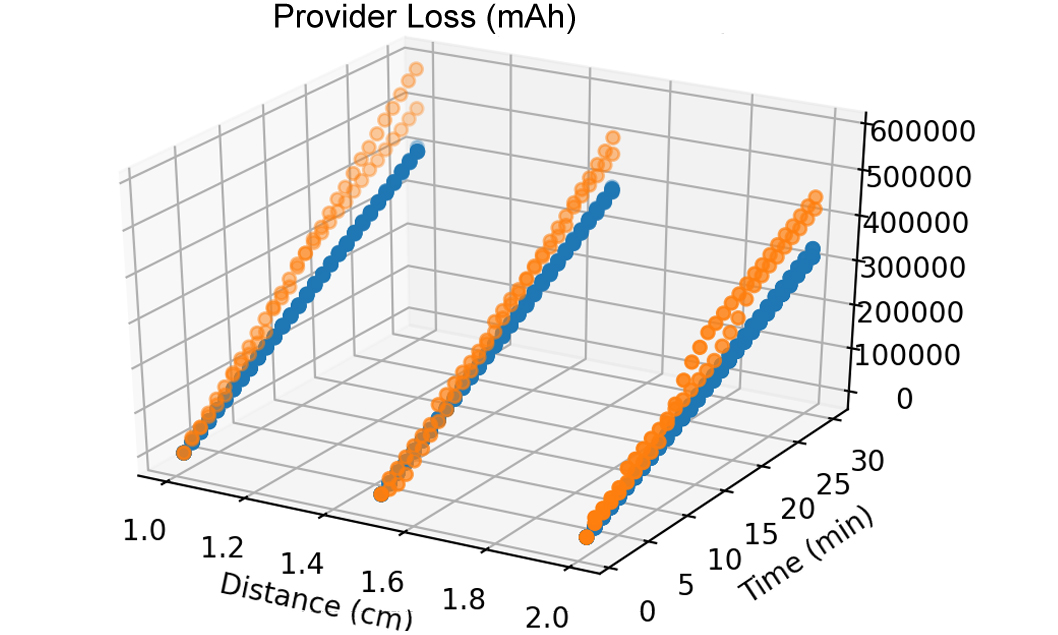}   
    \caption{Provider loss data (orange dots stand for outliers)}        
    \label{fig:datasets_provider}
\end{figure}

\begin{figure}[!t]
    \centering
     \setlength{\abovecaptionskip}{-2pt}
    \setlength{\belowcaptionskip}{-25pt}
        \includegraphics[width=0.86\linewidth]{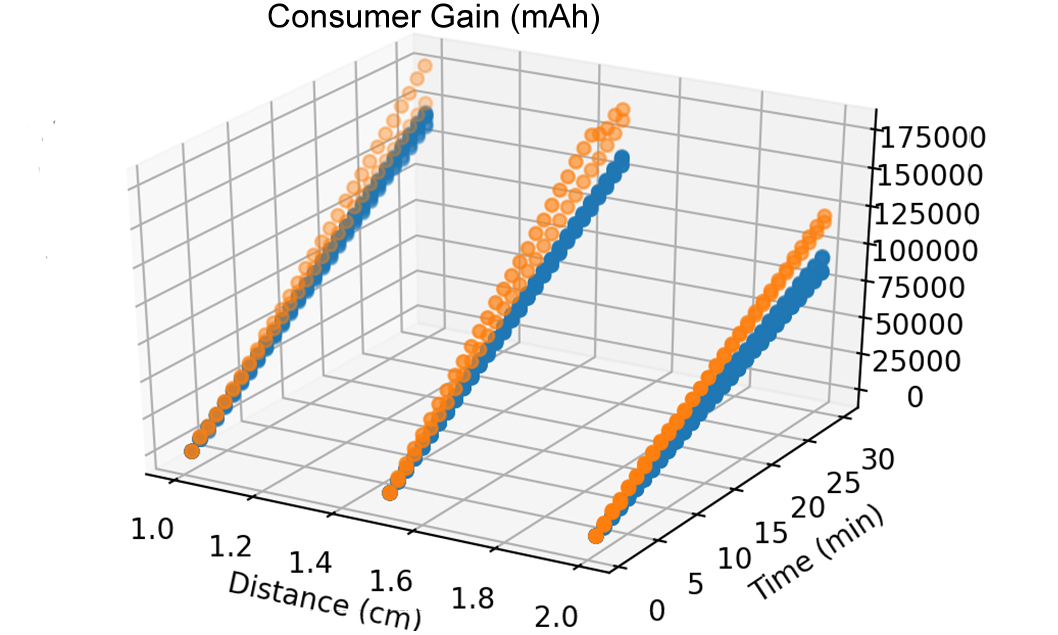}   
    \caption{Consumer gain data (orange dots stand for outliers)}        
    \label{fig:datasets_consumer}
\end{figure}

For the wireless energy sharing data collection experiments, we collected datasets over different wireless energy sharing distances, i.e.,  1 cm, 1.5 cm, and 2 cm. In order to simulate the IoT user preference for energy sharing, we repeated the experiments on the aforementioned distances 7, 14, and 21 times respectively. Therefore, we have 1260 data, i.e., (7+14+21)$\times$30, for the provider and consumer, respectively. Thus, we collected a total of 2520 data points. In summary, the dataset consists of three attributes: (1) the distance between the coils, (2) the charging duration, and (3) the battery levels of the consumer and provider at every $mt$ time interval in mAh. In our experiments, $mt$ was 1 minute. For the usage experiment, we conducted the experiment five times and collected 300 data records in total. The statistics for the experimental environment are listed in Table \ref{ExpVariables}. The datasets are shown in Fig.\ref{fig:datasets_provider} and Fig.\ref{fig:datasets_consumer}.\looseness=-1


\begin{table}[!t]
\renewcommand\arraystretch{1.1}
\footnotesize
\centering
\caption{Statistics of the experiment's environment}
\label{ExpVariables}
\setlength{\abovecaptionskip}{0pt}
\setlength{\belowcaptionskip}{0pt}
\begin{tabular}{l|l}
\hline
Variables & Value   \\ \hline
Duration of Charging            & 30 minutes   \\ 
Distances of Charging                  & 1 cm, 1.5 cm, 2 cm              \\ 
Number of Experiments on Above Distances                  & 7, 14, 21              \\ 
Data Amount on Above Distances                  & 420, 840, 1260              \\ 
Monitoring Interval $mt$                & 1 minute       \\ 
Experiment surrounding & 25 degrees Celsius  \\
Total Collected Data Amount            & 2820 \\ 
Energy Sharing Data Amount     & 2520  \\
Self-consumption Data Amount                & 300      \\
Preprocessed Data Amount                 & 2160      \\
Abnormal Charging Data Amount                 & 360      \\
  \hline
\end{tabular}
\vspace{-.3cm}
\end{table}

 \begin{figure}[!t]
    \centering
     \setlength{\abovecaptionskip}{-2pt}
    \setlength{\belowcaptionskip}{-25pt}
        \includegraphics[width=0.82\linewidth]{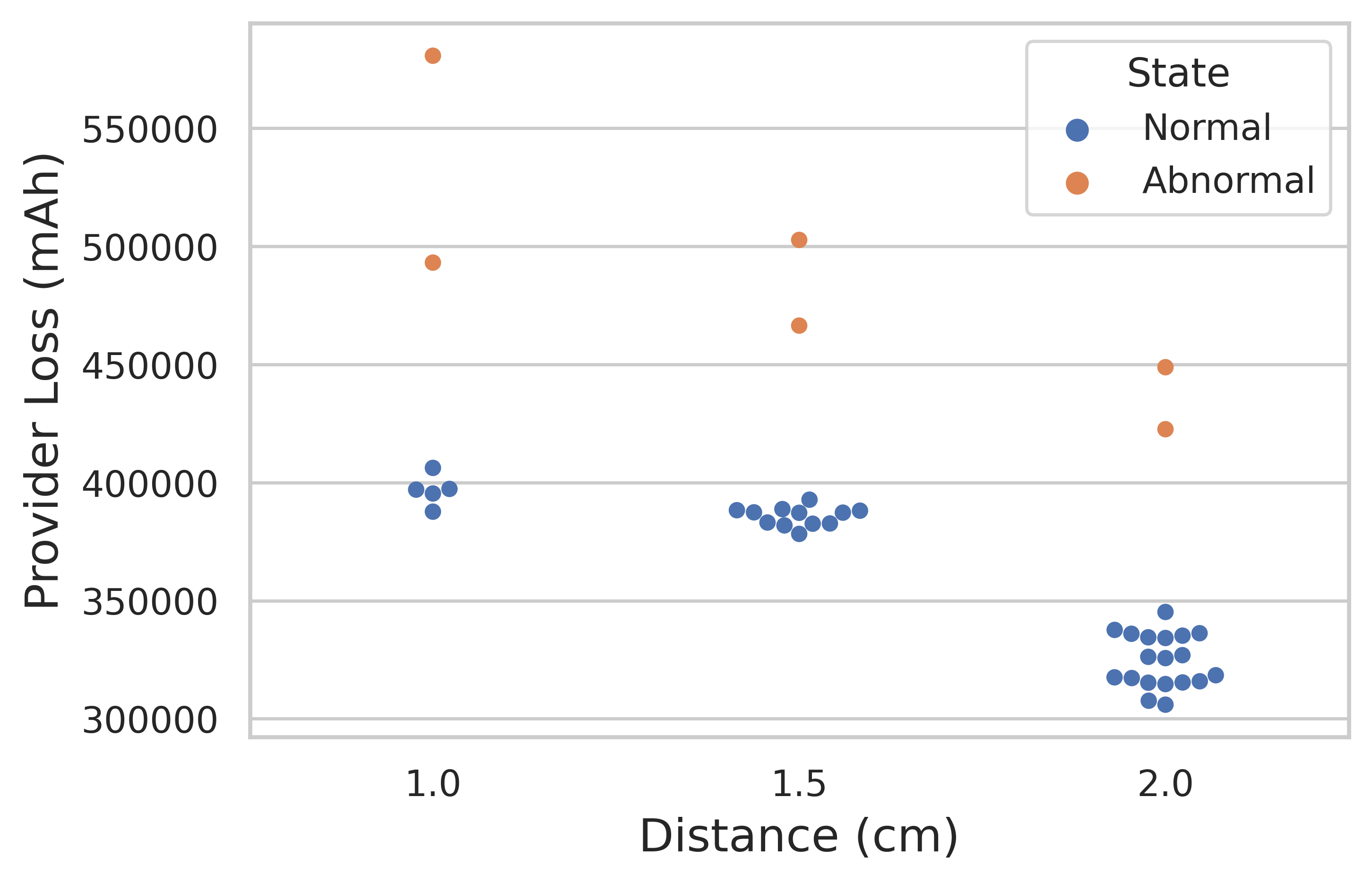}   
    \caption{Outlier detection for provider loss}        
    \label{fig:dbscan_provider}
\end{figure}

 \begin{figure}[!t]
    \centering
     \setlength{\abovecaptionskip}{-2pt}
    \setlength{\belowcaptionskip}{-25pt}
        \includegraphics[width=0.82\linewidth]{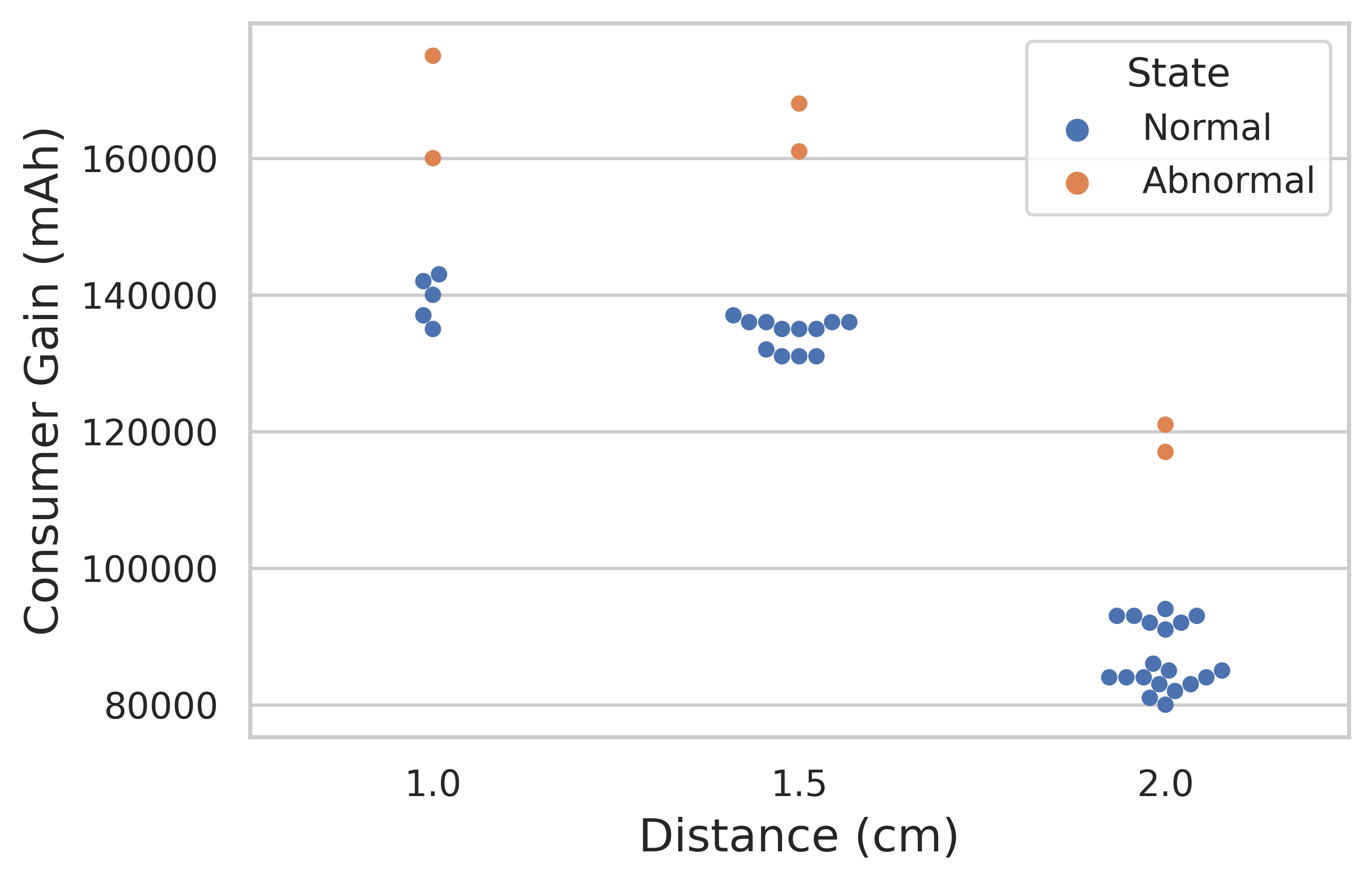}   
    \caption{Outlier detection for consumer gain}        
    \label{fig:dbscan_consumer}
\end{figure}
\subsection{Dataset Preprocessing}

The filtering phase of the framework can be viewed as a data preprocessing step (See Fig.\ref{fig:systemmodel}). Recall that the filter phase aims to eliminate outliers from the battery levels of consumers and providers in the energy sharing state. The performance of the filter on the battery data of the provider is shown in Fig.\ref{fig:datasets_provider} and Fig.\ref{fig:dbscan_provider}.  In these figures, the orange dots represent outliers. Similarly, the performance of the filter for the consumer battery data is shown in Fig.\ref{fig:datasets_consumer} and Fig.\ref{fig:dbscan_consumer}.  As stated in Section \ref{section}, the filtering phase focuses solely on the final status of the provider and consumer, i.e., $PL_{t_{k}}$ and $CG_{t_{k}}$, respectively. From Fig.\ref{fig:dbscan_provider} and Fig.\ref{fig:dbscan_consumer}, we can observe that the outliers are accurately detected. Specifically, we detected six abnormal charging data points comprising 360 data points (See Table \ref{ExpVariables}). We then eliminated all the detected outliers, after which we conducted the following experiments mainly based on datasets collected from normal energy sharing processes (blue dots in Fig.\ref{fig:datasets_provider} and Fig.\ref{fig:datasets_consumer}). After removing the outliers,  we used 2160 data points as input for the Easeformer. The data points consist of 1080 data points of provider loss $\mathcal{PL}$ and 1080 data points of consumer gain $\mathcal{CG}$ (See Table \ref{ExpVariables}).


\subsection{Evaluation of the ELP Framework}
In this section, we evaluate the effectiveness of our proposed ELP framework. Specifically, we conducted a set of experiments to evaluate the performance of the prediction and estimation phases (See Fig.\ref{fig:systemmodel}). We used the following settings to analyze the performance of our proposed ELP framework:
\begin{itemize}[ noitemsep,nosep,leftmargin=8pt,labelsep=4pt,itemindent=2pt]
\item \textbf{Metrics:} We use MSE and MAE as performance metrics, which can be respectively computed as follows:
     \begin{equation}
     \begin{aligned}
        \text{MSE} = \frac{1}{n}\sum^{n}_{i=1}(\hat{y_i}-y_i)^2 \;\;\;\;
\text{MAE} = \frac{1}{n}\sum^{n}_{i=1}\lvert  \hat{y_i}-y_i \lvert 
    \end{aligned}
    \vspace{-5pt}
     \end{equation}
where $\hat{y_i}$ represents the predictions and $y_i$ represents the ground truth.

\item \textbf{Baseline:} We compared our proposed Easeformer with the standard Informer using collected datasets. The Informer is a multi-head attention model that uses the ProbSparse attention and self-attention distilling technique to enhance its efficiency \cite{zhou2021informer}. 
\end{itemize}

 \begin{figure}[!t]
    \centering
     \setlength{\abovecaptionskip}{-2pt}
    \setlength{\belowcaptionskip}{-25pt}
        \includegraphics[width=0.69\linewidth,center]{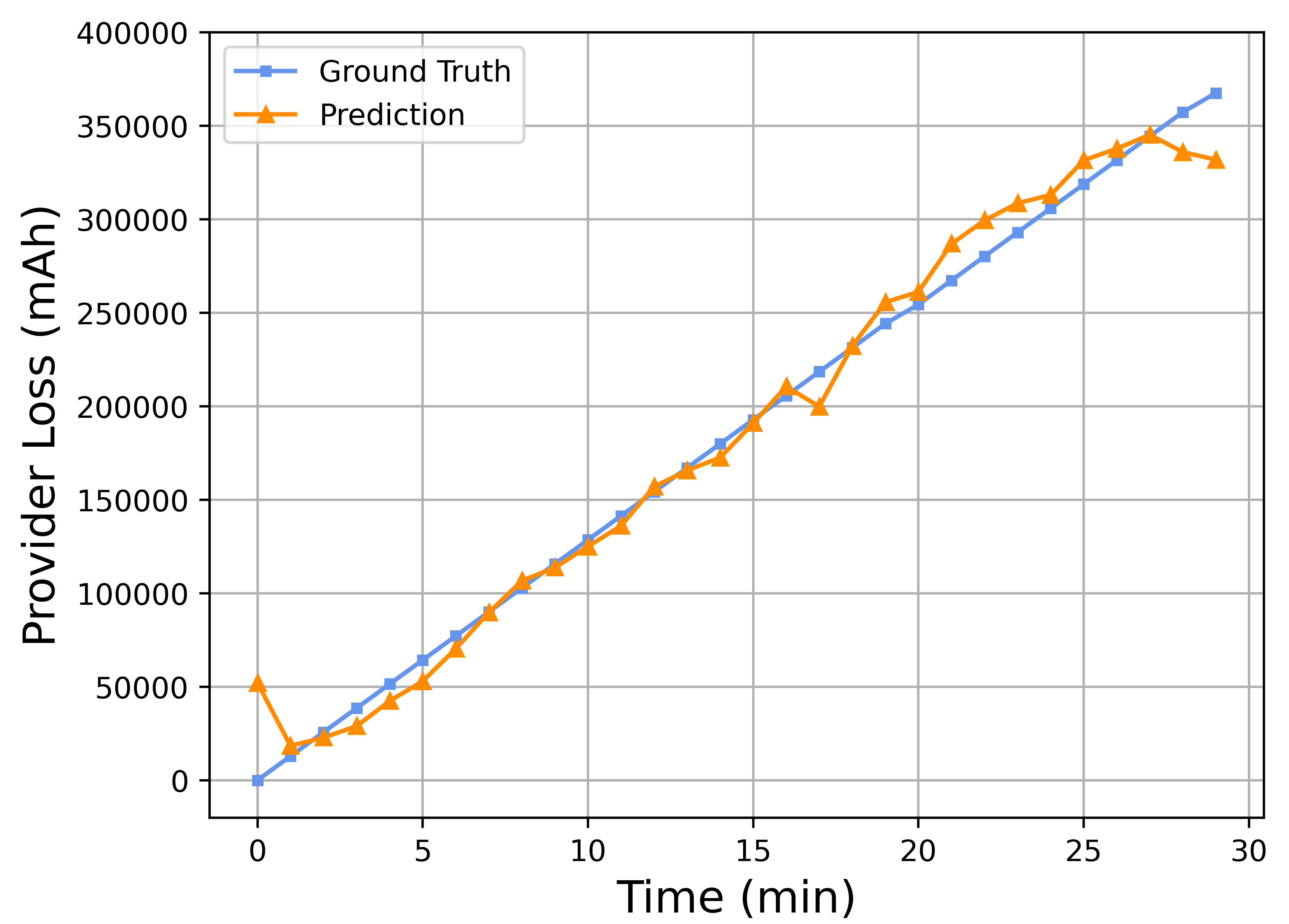}   
    \caption{Provider loss prediction}        
    \label{fig:eaas_provider}
\end{figure}

 \begin{figure}[!t]
    \centering
     \setlength{\abovecaptionskip}{-2pt}
    \setlength{\belowcaptionskip}{-25pt}
        \includegraphics[width=0.69\linewidth,center]{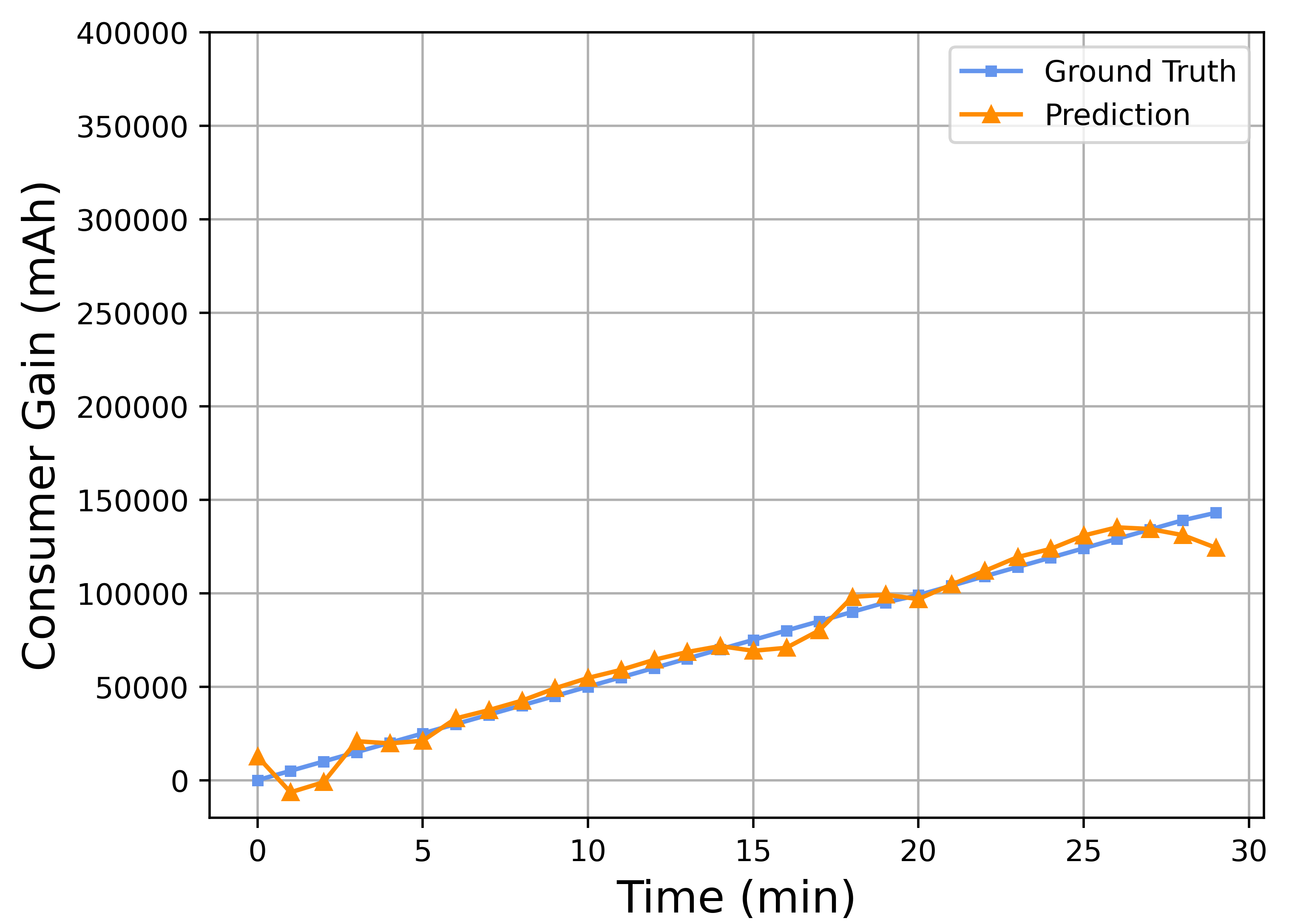}   
    \caption{Consumer gain prediction}        
    \label{fig:eaas_consumer}
\end{figure}




\begin{table*}[]
\renewcommand\arraystretch{1.36}
\centering

\small
\caption{Comparison and ablation experiment results of Easeformer}
\label{comparisonablation}
\begin{adjustbox}{width=0.76\textwidth}
\begin{tabular}{cccccccc}
\specialrule{.1em}{.05em}{.05em}
\multicolumn{2}{c}{\textbf{Generative Inference Length}}                                                                              & \multicolumn{2}{c}{\textbf{$L_{token}=$30}}                                                                                                                & \multicolumn{2}{c}{\textbf{$L_{token}=$60}}                                                                                                                & \multicolumn{2}{c}{\textbf{$L_{token}=$90}}                                                                       \\ \specialrule{.1em}{.05em}{.05em}
\multicolumn{1}{c|}{\textbf{Model Name}} & \multicolumn{1}{c|}{\textbf{Data Type}}                                                    & \textbf{MSE}                                                                & \multicolumn{1}{c|}{\textbf{MAE}}                                            & \textbf{MSE}                                                                & \multicolumn{1}{c|}{\textbf{MAE}}                                            & \textbf{MSE}                                            & \textbf{MAE}                                            \\ \specialrule{.1em}{.05em}{.05em}
\multicolumn{1}{c|}{Informer}            & \multicolumn{1}{c|}{\begin{tabular}[c]{@{}c@{}}Provider Loss\\ Consumer Gain\end{tabular}} & \multicolumn{1}{l}{\begin{tabular}[c]{@{}l@{}}0.1199\\ 0.1329\end{tabular}} & \multicolumn{1}{l|}{\begin{tabular}[c]{@{}l@{}}0.1984\\ 0.2147\end{tabular}} & \multicolumn{1}{l}{\begin{tabular}[c]{@{}l@{}}0.1192\\ 0.1419\end{tabular}} & \multicolumn{1}{l|}{\begin{tabular}[c]{@{}l@{}}0.1870\\ 0.2238\end{tabular}} & \begin{tabular}[c]{@{}c@{}}0.0898\\ 0.1079\end{tabular} & \begin{tabular}[c]{@{}c@{}}0.1580\\ 0.1893\end{tabular} \\ \hline
\multicolumn{1}{c|}{Easeformer$^\dag$}   & \multicolumn{1}{c|}{\begin{tabular}[c]{@{}c@{}}Provider Loss\\ Consumer Gain\end{tabular}} & \begin{tabular}[c]{@{}c@{}}0.1262\\ 0.1372\end{tabular}                     & \multicolumn{1}{c|}{\begin{tabular}[c]{@{}c@{}}0.2060\\ 0.2266\end{tabular}} & \begin{tabular}[c]{@{}c@{}}0.1176\\ 0.1255\end{tabular}                     & \multicolumn{1}{c|}{\begin{tabular}[c]{@{}c@{}}0.1968\\ 0.2161\end{tabular}} & \begin{tabular}[c]{@{}c@{}}\textbf{0.0765}\\ \textbf{0.0980}\end{tabular} & \begin{tabular}[c]{@{}c@{}}\textbf{0.1493}\\ 0.1743\end{tabular} \\ \hline
\multicolumn{1}{c|}{Easeformer$^\ddag$}   & \multicolumn{1}{c|}{\begin{tabular}[c]{@{}c@{}}Provider Loss\\ Consumer Gain\end{tabular}} & \begin{tabular}[c]{@{}c@{}}\textbf{0.1104}\\ \textbf{0.1262}\end{tabular}                     & \multicolumn{1}{c|}{\begin{tabular}[c]{@{}c@{}}\textbf{0.1711}\\ \textbf{0.1893}\end{tabular}} & \begin{tabular}[c]{@{}c@{}}\textbf{0.1028}\\ \textbf{0.1135}\end{tabular}                     & \multicolumn{1}{c|}{\begin{tabular}[c]{@{}c@{}}\textbf{0.1659}\\ \textbf{0.1854}\end{tabular}} & \begin{tabular}[c]{@{}c@{}}0.0834\\ 0.1002\end{tabular} & \begin{tabular}[c]{@{}c@{}}0.1526\\ \textbf{0.1707}\end{tabular} \\ \specialrule{.1em}{.05em}{.05em}
\multicolumn{8}{l}{\begin{tabular}[c]{@{}l@{}}$^1$ Easeformer$^\dag$ removes the encoder input transformer mechanism\\ $^2$ Easeformer$^\ddag$ uses the encoder input transformer mechanism\end{tabular}}                                                          
\end{tabular}
\end{adjustbox} \vspace{-13pt}
\end{table*}

\begin{table*}[]
\renewcommand\arraystretch{1.26}
\centering
\small
\caption{Experiment results of linear regression models and ELP}
\label{linearelp}
\begin{adjustbox}{width=0.76\textwidth}
\begin{tabular}{cccc}
\specialrule{.1em}{.05em}{.05em}
\multicolumn{2}{c}{\textbf{Metrics}}                                                                                                                            & \textbf{MSE}                                                                                           & \textbf{MAE}                                                                      \\ \specialrule{.1em}{.05em}{.05em}
\multicolumn{1}{c|}{\textbf{Model Name}}   & \multicolumn{1}{c|}{\textbf{Data Type}}                                                                            & \multicolumn{1}{c|}{\textbf{Mean  \:\:  Std\:}}                                                              & \textbf{Mean  \:\:  Std\:}                                                              \\ \specialrule{.1em}{.05em}{.05em}
\multicolumn{1}{c|}{Linear Regression}     & \multicolumn{1}{c|}{\begin{tabular}[c]{@{}c@{}}Provider Self-consumption\\ Consumer Self-consumption\end{tabular}} & \multicolumn{1}{c|}{\begin{tabular}[c]{@{}c@{}}0.00057    0.00023 \\  0.00091    0.00029\end{tabular}} & \begin{tabular}[c]{@{}c@{}}0.01852    0.00298 \\  0.03542    0.00863\end{tabular} \\ \hline
\multicolumn{1}{c|}{Energy Loss Prediction} & \multicolumn{1}{c|}{Entire ELP Framework}                                                                          & \multicolumn{1}{c|}{0.11676    0.02591}                                                                & 0.26784    0.01917                                                                \\ \specialrule{.1em}{.05em}{.05em}
\end{tabular}
\end{adjustbox}
\vspace{-.38cm}
\end{table*}

 \begin{figure}[!t]
    \centering
     \setlength{\abovecaptionskip}{-2pt}
    \setlength{\belowcaptionskip}{-30pt}
        \includegraphics[width=0.8\linewidth, center]{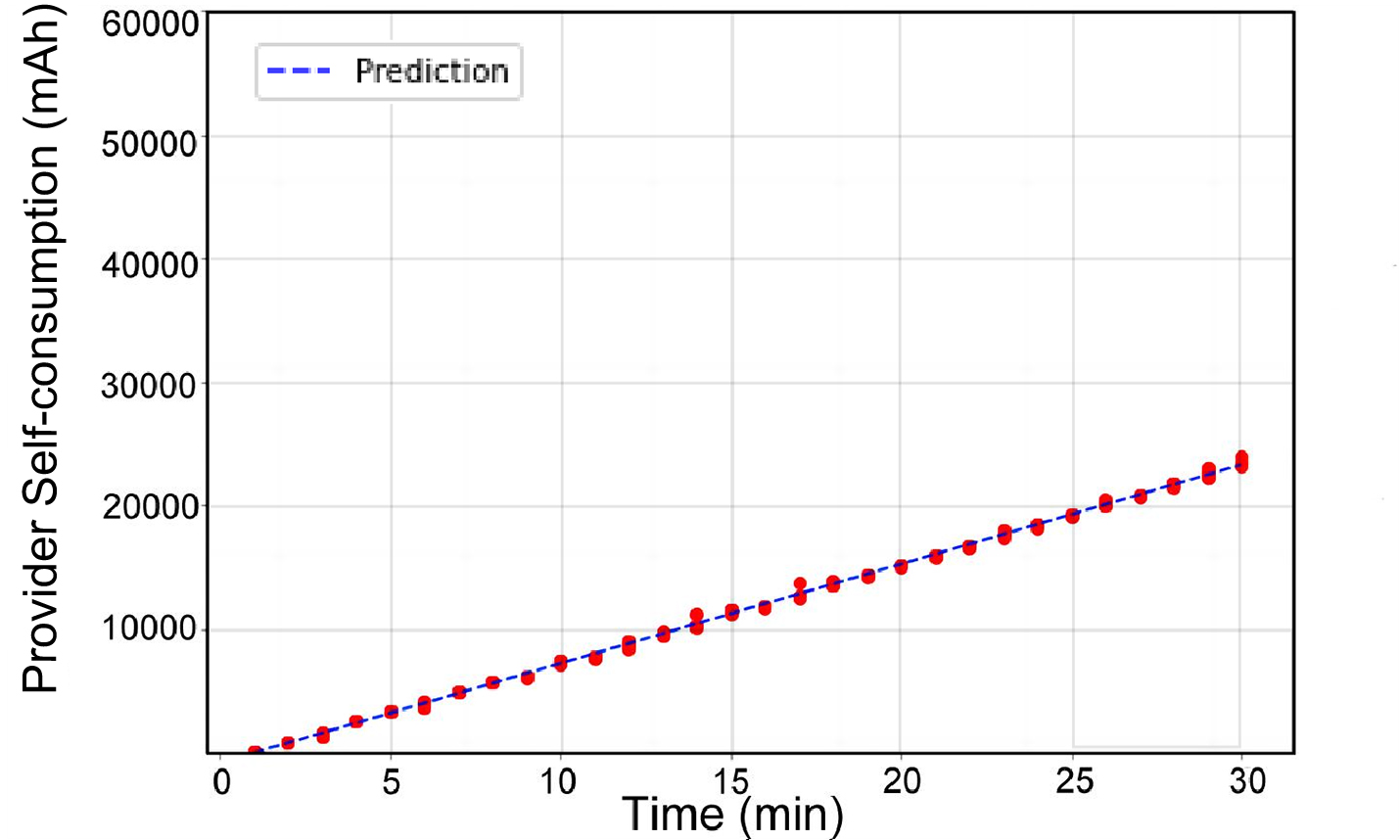}   
    \caption{Provider self-consumption (usage) prediction}       \vspace{-10pt} 
    \label{fig:provider_linear}
\end{figure}

 \begin{figure}[!t]
    \centering
     \setlength{\abovecaptionskip}{-2pt}
    \setlength{\belowcaptionskip}{-25pt}
        \includegraphics[width=0.8\linewidth, center]{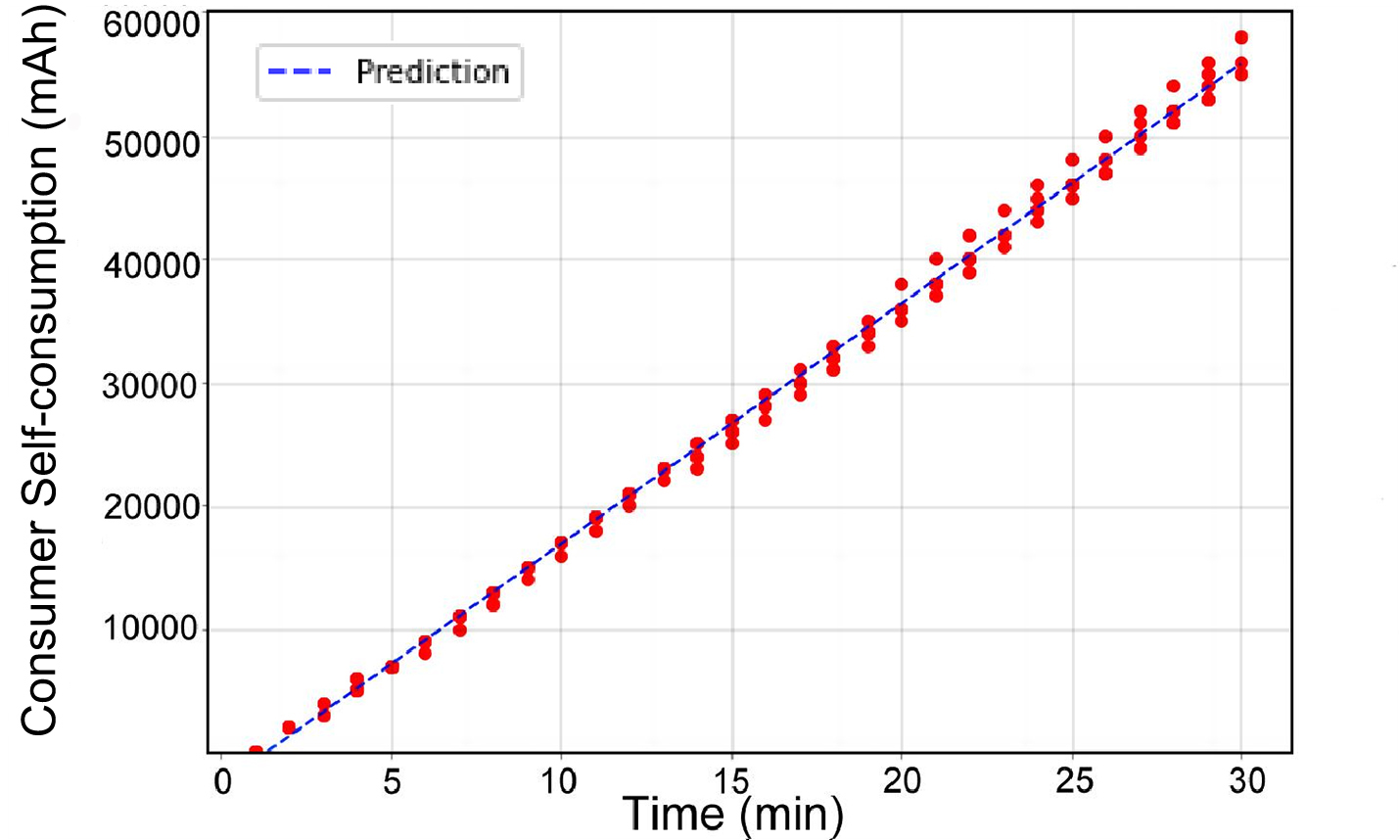}   
    \caption{Consumer self-consumption (usage) prediction}        
    \label{fig:consumer_linear}
\end{figure}

In the prediction phase, we first conducted a series of comparison experiments to demonstrate that our Easeformer surpasses the state-of-the-art algorithm in handling energy-sharing datasets. We selected a time series forecasting method, Informer, as our baseline. Next, we performed a grid search over the hyperparameters. Easeformer was optimized using the Adam optimizer, with an initial learning rate of 1$e^{-4}$, decaying by a factor of 0.5 every epoch. The total number of epochs was set to 10 and early stopping was implemented as needed. The batch size was set to 8. The training, validation, and test datasets comprised 50\%, 25\%, and 25\% of the dataset, respectively. The input of each dataset was zero-mean normalized, except for feature distance $\mathcal{D}$. The softmax temperature in EIT was set to 0.85. The head number of the multi-head attention mechanism $h$ (See Fig.\ref{Easeformer}) was set to 8. Furthermore, as the length of the start token, i.e., $L_{token}$, plays a crucial role in partial generative inference and can influence the performance of Easeformer, we carry out comprehensive experiments to demonstrate the impact of $L_{token}$ on the effectiveness of Easeformer. Specifically, we progressively extended the length of the start token $L_{token}$, i.e., $L_{token}$ $=$ $30$, $L_{token}$ $=$ $60$, $L_{token}$ $=$ $90$. Figures \ref{fig:eaas_provider} and \ref{fig:eaas_consumer} separately display the provider loss and consumer gain prediction when the energy sharing distance is equal to 1.5 cm and $L_{token}$ is set to 90. Table \ref{comparisonablation} presents the results of the comparison experiments between the Informer and Easeformer under varying start token lengths. The best outcomes for each start token length are emphasized in bold font. From Table \ref{comparisonablation}, we observe that the proposed Easeformer significantly enhances the prediction effectiveness, demonstrating its success in improving the predictive capacity for energy sharing data. We also conducted supplementary ablation experiments to illustrate the effectiveness of our encoder input transformer EIT mechanism (See Algorithm \ref{alg:one}). In the ablation experiment, we used Easeformer$^\ddag$ as a benchmark to eliminate the additional effects of EIT. According to Table \ref{comparisonablation}, Easeformer$^\ddag$ completes all the experiments and achieves superior performance, particularly when the start token length is short, i.e., $L_{token}$ $=$ $30$ and $L_{token}$ $=$ $60$. The comparison method, Easeformer$^\dag$ omits EIT. Considering the advantages of incorporating user distance preference probability as input for the encoder in the energy sharing prediction problem, we conclude that adopting EIT is worthwhile, particularly when distance serves as a crucial indicator and $L_{token}$ is set to a small value.\looseness=-1

\vspace{0pt}
Additionally, we evaluated the performance of linear regression models. Figures \ref{fig:provider_linear} and \ref{fig:consumer_linear} illustrate the effectiveness of linear regression models. We observe that the self-consumption of both the provider and consumer is stable over time, and the linear regression model is proficient at fitting usage (self-consumption) data. The performance of the linear regression models is shown in Table \ref{linearelp}.\looseness=-1

\begin{figure}[!t]
    \centering
     \setlength{\abovecaptionskip}{-2pt}
        \includegraphics[width=0.75\linewidth,center]{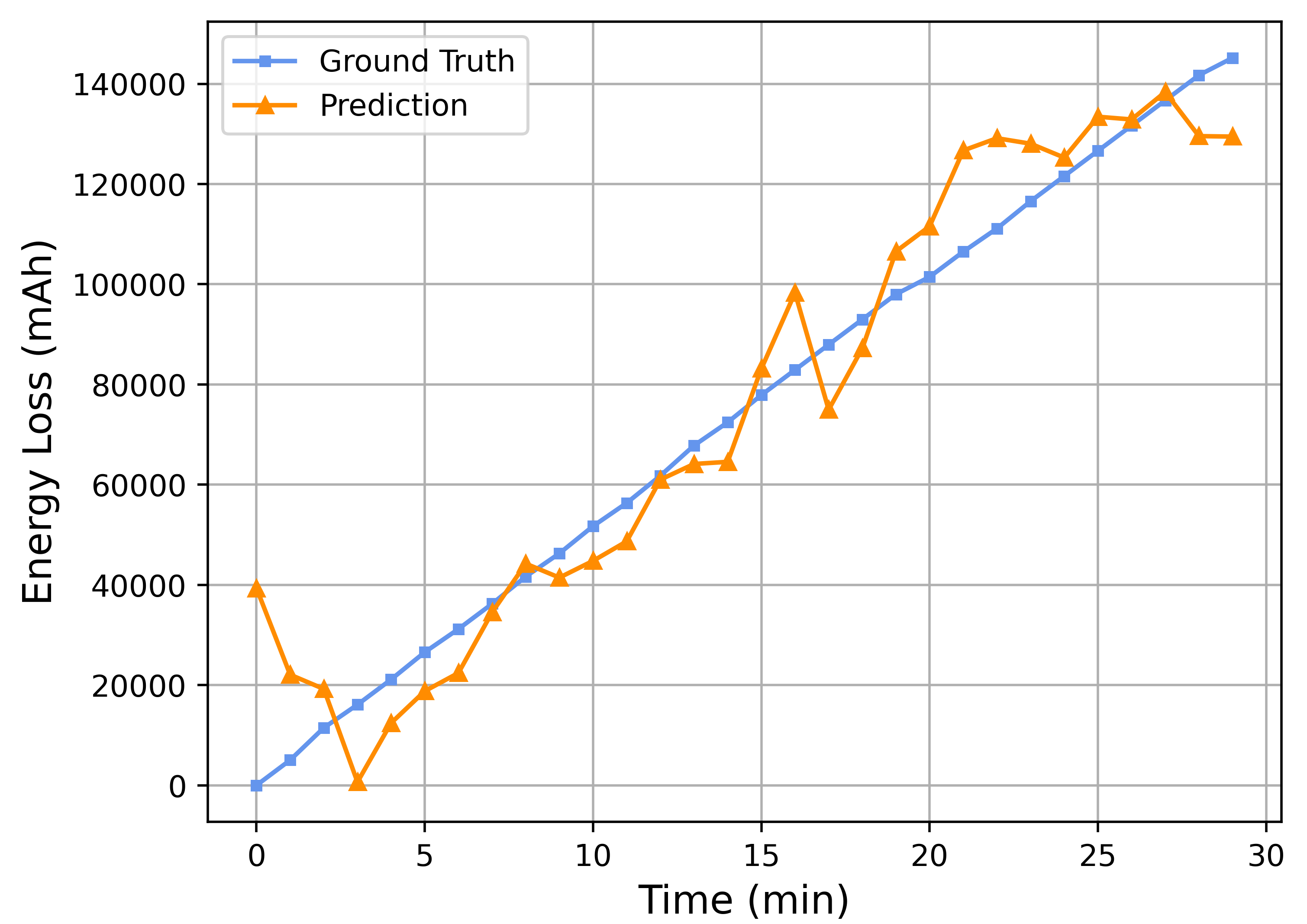}   
    \caption{Energy loss prediction}        
    \label{fig:energyloss}
\end{figure}

In the final experiment, we assessed the effectiveness of the ELP framework. The ELP framework aims to predict the energy loss derived from the energy sharing process. The performance of the ELP framework is displayed in Table \ref{linearelp}. Because the energy loss was computed based on the outputs of the two Easeformer and two linear regression models, the MSE and MAE were higher than the individual results of the aforementioned models. Figure \ref{fig:energyloss} illustrates the effectiveness of the ELP framework. From Fig.\ref{fig:energyloss}, we observe that our ELP framework is adept at fitting the ground truth of energy loss.



\section{Conclusion}
In this paper, we proposed an Energy Loss Prediction (\textit{ELP}) framework for estimating the energy loss during the delivery of wireless energy services. The framework employs two models to predict the battery levels for the consumer and the provider. We introduced an Easeformer to predict the battery levels of IoT devices during the energy sharing state. Additionally, we designed an Encoder Input Transformer (\textit{EIT}) for the EaseFormer, which dynamically considers the impact of the wireless energy sharing distance. Moreover, we proposed a decoder that features a partial generative inference. Experiments conducted on real-world data demonstrated the effectiveness of the Energy Loss Prediction framework and highlighted the superior performance of the Easeformer compared to the Informer in the context of energy sharing problems.\looseness=-1
\vspace{-5pt}
\section*{Acknowledgment} 
This research was partly made possible by LE220100078 and DP220101823 grants from the Australian Research Council. The statements made herein are solely the responsibility of the authors.

\balance
\bibliographystyle{IEEEtran}
\bibliography{main}

\end{document}